\documentclass[
a4paper,floatfix,
amsmath,amssymb,
aps,
pre,
reprint,
]{revtex4-1}

\usepackage{fancyhdr}
\usepackage{graphicx}
\usepackage{epsfig}
\usepackage{dcolumn}
\usepackage{bm}
\usepackage[mathlines]{lineno}
\usepackage{amsmath,esint}
\usepackage{color}
\usepackage[usenames,dvipsnames,svgnames,table]{xcolor}
\usepackage{url}
\usepackage{mathtools}
\usepackage{esvect}
\usepackage{enumitem}
\usepackage[caption=false]{subfig}
\usepackage{float}
\usepackage{soul}  % highlight

\newcommand{\grad}{\vec{\nabla}}
\newcommand{\vecb}[1]{\mathbf{#1}}
\newcommand{\dt}[1]{\frac{\partial #1}{\partial t}}
\newcommand\lvec[1]{\mkern1.5mu\vv{\mkern-1.5mu#1}}

\newcommand{\period}{\text{.}} 
\newcommand{\comma}{\text{,}} 

\newcommand{\supers}[1]{\textsuperscript{#1}}
\newcommand{\na}{---}

\usepackage{natbib,hyperref}
\hypersetup{
     colorlinks   = true,
     citecolor    = blue,
     linkcolor    = blue,
     urlcolor     = blue
}

\begin{document}
	
\title{Dynamic coupling between particle-in-cell and atomistic simulations}

\author{Mihkel Veske}
\email{mihkel.veske@helsinki.fi}
\author{Andreas Kyritsakis}
\author{Flyura Djurabekova}
\affiliation{Department of Physics and Helsinki Institute of Physics, University of Helsinki, P.O. Box 43 (Pietari Kalmin katu 2), 00014 Helsinki, Finland}

\author{Kyrre Ness Sjobak}
\affiliation{CERN, Geneva, Switzerland}
\affiliation{Department of Physics, University of Oslo, P.O. Box 1048 Blindern, N-0316 Oslo, Norway}

\author{Alvo Aabloo}
\author{Vahur Zadin}
\affiliation{Intelligent Materials and Systems Lab, Institute of Technology, University of Tartu, Nooruse 1, 50411 Tartu, Estonia}

\date{18.05.2020}
\renewcommand{\headrulewidth}{0pt}
\renewcommand{\footrulewidth}{0.5pt}
\lhead{}
\rhead{}
\lfoot{Veske \textit{et al.}}
\rfoot{\thepage}
\cfoot{\href{https://doi.org/10.1103/PhysRevE.101.053307}{Phys. Rev. E, 101(5), p.053307}}

\begin{abstract}
    We propose a method to directly couple molecular dynamics, the finite element method, and particle-in-cell techniques to simulate metal surface response to high electric fields.
    We use this method to simulate the evolution of a field-emitting tip under thermal runaway by fully including the three-dimensional space-charge effects.
    We also present a comparison of the runaway process between two tip geometries of different widths.
    The results show with high statistical significance that in the case of sufficiently narrow field emitters, the thermal runaway occurs in cycles where intensive neutral evaporation alternates with cooling periods.
    The comparison with previous works shows, that the evaporation rate in the regime of intensive evaporation is sufficient to ignite a plasma arc above the simulated field emitters.
    
    \vspace{5pt}
    Keywords: vacuum arc, vacuum breakdown, electron emission, plasma initiation, thermal runaway, metal nanotip, multiscale-multiphysics simulation
\end{abstract}

\maketitle

\section{Introduction}
\label{sec:introduction}

Field-emitting tips play a detrimental role in various vacuum devices that require high electric fields, such as vacuum interrupters \cite{slade2007}, x-ray sources \cite{knaw_smart*light:_2019}, fusion reactors \cite{McCracken1980}, and particle accelerators \cite{cern_compact_2018}.
When a field emitter is subjected to currents exceeding a certain threshold, the emission becomes unstable and swiftly results in a vacuum arc \cite{Dyke1953Arc, Dyke1953I}.
The latter is characterized by the ignition of a plasma in the vacuum gap, which drives a large current and converts the gap into a short-circuit.
The ignition of an arc in vacuum (vacuum breakdown) can cause catastrophic failure in electron sources \cite{Anders_PRL88,Dyke1953Arc}.

Although vacuum arcs have been studied more than half a century and it is well-known that they appear after intense field electron emission \cite{Dyke1953I,Anders}, the physical mechanisms that lead from field emission to plasma ignition without the presence of any ionizable gas are not understood yet.
One hypothesis commonly used to explain the plasma build-up in vacuum after intense field emission is the ``explosive emission'' scenario \citep{Anders,mesyats1993ectons,Mesyats2005}.
According to it, when intensive field emission takes place, there is a critical current density beyond which heating is produced at a rate the emitter cannot dissipate, which leads to heat accumulation, sufficient to cause instant explosion and plasma formation.
This phenomenological description does not give an insightful physical understanding of the underlying processes and cannot \textit{a priori} predict the behavior of a specific system.
Such empirical considerations have been common in the vacuum arcing theory since the 1950-s, as the modern computational tools necessary to describe the physical processes of such a complex phenomenon as vacuum arcing were not available.
Thus, there is a growing need for the development of theoretical concepts and computational tools that describe the vacuum arc ignition in a rigorous quantitative manner and provide a deep understanding of the physical processes that lead to it.

This has changed recently, since various computational methods have been employed to study vacuum arcs.
In a recent work \cite{timko_field_2015}, plasma formation in Cu cathodes was studied by means of particle-in-cell simulations.
It was found that plasma can build up in the vicinity of an intensively field-emitting cathode, assuming that not only electrons, but also neutral atoms are emitted from the cathode surface. 
Moreover, it was shown that it is sufficient to supply about 0.015 neutrals per electron to initiate the ionization avalanche leading to the formation of a stable plasma.
Recent multiscale atomistic simulations on Cu nanotips \cite{kyritsakis_thermal_2018} revealed a thermal runaway process which causes field-assisted evaporation of Cu atoms and nanoclusters at a rate exceeding the threshold found in \cite{timko_field_2015}.
Thus, combining these two independently obtained results, one can consider intensively field-emitting tips as plausible sources of plasma, since they can supply both species, electrons and neutrals, at individual rates.

However, it is not clear yet whether a field-emitting tip is indeed able to produce sufficient densities to initiate the ionization avalanche, since there is no technique which can address this issue concurrently.
To answer this question, a fully coupled molecular dynamics (MD) and particle-in-cell (PIC) simulation technique must be developed. 
The MD part will describe the behavior of atoms under given conditions, such as heat generated by field emission currents and stress exerted by the field, while the PIC part will follow the dynamics of the particles above the surface.
In this way, the space-charge buildup above the field-emitting tip can also be found directly, without a need for simplifying assumptions.

In \cite{kyritsakis_thermal_2018}, we attempted to take into account the space-charge effect by using a semiempirical approximate model, as proposed in \cite{forbes_exact_2008}. 
This model treats the emitter as planar, using a universal correction factor to account for the three-dimensional (3D) nature of the geometry, which was an external adjustable parameter for our simulations, with its value roughly approximated based on previous PIC calculations \cite{Uimanov2011}. 
Although we obtained promising results with this model, the semiempirical correction does not capture the complex 3D distribution of the space-charge developed around the dynamically varying tip geometry.
Moreover, the accuracy of the calculations of the space-charge affects directly the value of the field emission currents and all the subsequent processes developing in the tip due to the field emission process.

In this paper, we present an approach of the coupled MD-PIC technique for a more accurate estimation of processes contributing to plasma onset.
In particular, we develop a method that directly couples PIC and MD simulations of the metal surface response to the field aided by the use of the finite element method (FEM).
We use the model to simulate similar process as in \cite{kyritsakis_thermal_2018}, but including accurately the 3D space-charge distribution that builds up above the highly curved nanometric field-emitting tip.
There are some attempts in the literature to simulate the evolution of the space-charge using an MD-like method \cite{torfason_molecular_2015,torfason_molecular_2016} where all interelectron interactions are considered.
For specific applications, such models may be sufficient, but for computational efficiency we prefer to use the PIC method.
The latter provides a sufficiently reliable kinetic description of space-charge or plasma evolution and allows incorporating the complex nature of the atomic processes and plasma-surface interactions, while maintaining a feasible computational load as compared to the MD-like methods~\cite{tskhakaya_particle--cell_2007,timko_field_2015}.
We also present a comparison of the runaway process between two tip geometries of different widths.
That comparison takes us closer to obtaining an estimation for the runaway occurrence probability in structures observed in experiments \cite{timko_modelling_2011}.
The incorporation of the PIC method into our simulations forms a decisive step towards the direct coupling between atomistic and plasma simulations, in order to fully reveal the processes that take place during the initiation of a vacuum arc.

\section{Method}
\label{sec:method}
\begin{figure}[h!]
    \centering
    \includegraphics[width=0.97\linewidth]{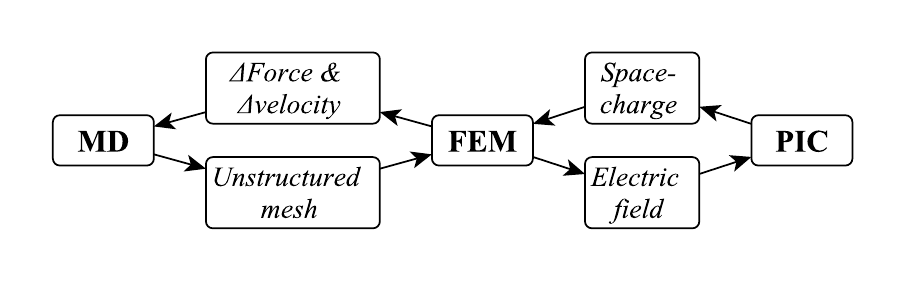}
    \vspace{-10pt}
    \caption{The relation between different components in the proposed multiscale-multiphysics model.}
    \label{fig:femocs_short}
\end{figure}

We present here a model that combines classical MD, FEM, PIC and electron emission calculations (fig.~\ref{fig:femocs_short}).
The model allows us to assess the effect of different physical processes on the dynamic evolution of the atomic structure of metal surfaces exposed to high electric fields. 
While the movement of the atoms is followed by an MD algorithm, the movement of the emitted electrons is tracked using the PIC method.
FEM provides the tools for concurrent and self-consistent electric field and heat diffusion calculations, whose output affects the force and velocity calculations in MD and the force exerted on the electrons in PIC.
Such a methodology forms the conceptual basis for the full coupling between metal surface and plasma simulations.
A full schematic of the model is provided in appendix \ref{sec:femocs_long}.

\vspace{-10pt}

The model is based upon the previously developed framework FEMOCS \cite{veske_dynamic_2018}.
In FEMOCS, an unstructured mesh is dynamically built around surface atoms, the density of which is controlled independently in all regions of the simulation domain.
The mesh consists of linear hexahedral elements that are built into tetrahedra that allow achieving satisfactory accuracy while keeping the computational cost within reasonable limits.
For more details about meshing, local solution extraction, and optimization features in FEMOCS, see \cite{veske_dynamic_2018}.
In the following sections we provide more details about the newly developed methodology for calculating electric field, electrostatic forces, and velocity perturbation.

\subsection{Electric field}

\begin{figure}[h!]
    \includegraphics[scale=1.0]{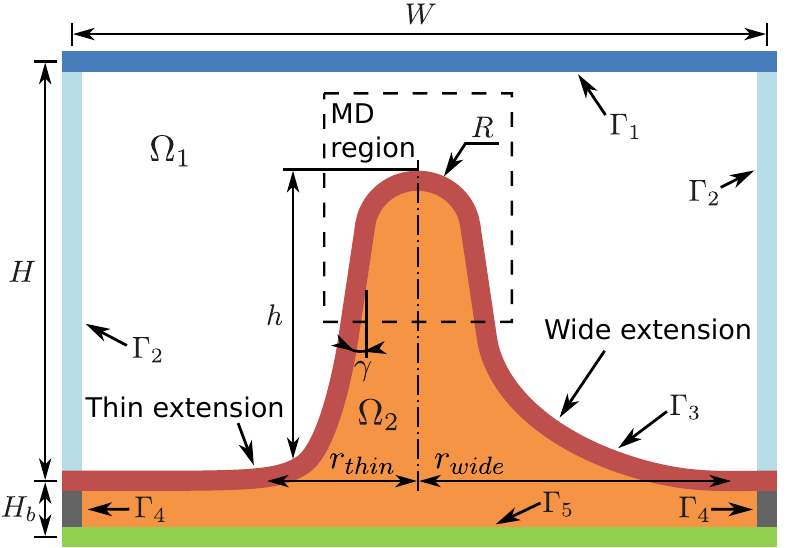}
    \caption{Boundary regions and geometry of the simulation domain, which consists of an atomistic region and an extension.}
    \label{fig:boundary}
\end{figure}

A common way to take into account an electric field within a computational model is to define it as the negative gradient of the electrostatic potential $\Phi$ that, in turn, can be calculated by solving the Poisson's equation in the vacuum (domain $\Omega_1$ in fig.~\ref{fig:boundary})
\begin{equation}
\label{eq:poisson}
\grad \cdot (\epsilon \grad \Phi) = -\rho \comma
\end{equation}
where $\epsilon$ is the dielectric constant and $\rho$ is the space-charge density.
We solve eq. \eqref{eq:poisson} iteratively with the boundary conditions (BC) defined in the vacuum (see fig.~\ref{fig:boundary}) as
\begin{equation}
\label{eq:poisson_bc}
\begin{aligned}
\grad \Phi \cdot \vec{n} = E_0 &\text{ on } \Gamma_1 \comma\\
\grad \Phi \cdot \vec{n} = 0 &\text{ on } \Gamma_2 \comma\\
\Phi = 0 &\text{ on } \Gamma_3 \comma
\end{aligned}
\end{equation}
where $E_0$ is the long-range applied electric field.
The boundary value problem \eqref{eq:poisson}-\eqref{eq:poisson_bc} is solved in FEMOCS by means of FEM. The derivation of weak and algebraic formulation of eq. \eqref{eq:poisson}-\eqref{eq:poisson_bc} is provided in appendix \ref{sec:weak_poisson}.

\subsection{PIC method}

To obtain the space-charge distribution $\rho(\vec{r})$ in the vacuum above the metal surface, which enters eq. \eqref{eq:poisson}, we use the PIC method \cite{birdsall_plasma_1985,buneman_dissipation_1959,dawson_one-dimensional_1962}.
Within PIC, the space-charge is formed by a set of superparticles (SPs), which is a common approach for space-charge and plasma simulations.
Each SP is a group of charges of the same type, electrons or ions, that act within the model as a single particle.
In other words, the SPs represent a sampling of the continuous phase space occupied by the real particles like electrons or ions.
Such a grouping is justified by the fact that the electrostatic and electrodynamic interactions depend on the charge-to-mass ratio instead of the charge or the mass of the particles.
The size of each SP is determined by its weight $w_{sp}$, which corresponds to the (fractional) number of particles it represents.
The SP weight is determined independently for each simulation, in order to achieve a sufficiently smooth charge distribution with a feasible number of SPs.

\begin{figure*}
    \includegraphics{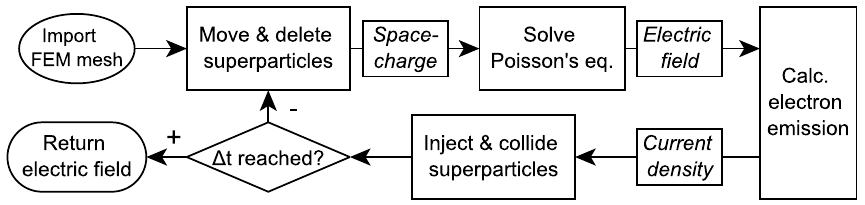}
    \caption{Electric field calculation by means of particle-in-cell method.}
    \label{fig:pic}
\end{figure*}

The movement of SPs is affected by the global electric field, which includes the long-range electrostatic interactions between particles.
The short-range interactions with nearby SPs are taken into account by means of binary Monte Carlo collisions (see appendix \ref{sec:collisions} for details).
This makes it possible to significantly enhance the computational efficiency and simulate large systems for long times.
Furthermore, the binary collision technique can be used to consider the interactions of different types of particle species in a plasma, such as impact ionizations.
Our current model, however, includes only the electrons, ignoring the presence of ions as well as electron-neutral interactions.
Within this approximation, we can study the processes at the early stage of plasma formation, which is the focus of the current work.
However, the model can be extended further if later stages of plasma evolution are of interest.
In the following four sections, we give an overview of the various stages of the PIC model, which are summarized in fig. \ref{fig:pic}.

\subsubsection{Particle mover}
\label{sec:particle_mover}
The propagation of SPs is performed by solving iteratively Newton's equations of motion for each SP:
\begin{equation}
\label{eq:newton}
\begin{aligned}
\dot{\vec{r}}&=\vec{v} \comma\\
\dot{\vec{v}}&=\frac{q}{m}\vec{E} \comma
\end{aligned}
\end{equation}
where $\frac{q}{m}$ is the charge-to-mass ratio of SPs.
In eq. \eqref{eq:newton}, SPs are affected only by the electrostatic field, since the magnetic field which is induced by the currents within the tip is negligible.
We solve eq. \eqref{eq:newton} numerically by using the leapfrog integration scheme \cite{birdsall_plasma_1985}.
In this method, the positions and the velocities of SPs are calculated as
\begin{subequations}
    \begin{align}
    \label{eq:leapfrog1}
    \vec{v}_{k+0.5} &= \vec{v}_k + Q\vec{E}_k \comma\\
    \label{eq:leapfrog2}
    \vec{r}_{k+1} &= \vec{r}_k + \Delta t \vec{v}_{k+0.5} \comma\\
    \label{eq:leapfrog3}
    \vec{v}_{k+1} &= \vec{v}_{k+0.5} + Q\vec{E}_{k+1} \comma
    \end{align}
\end{subequations}
where the subscript indicates the time step, $Q = \frac{q}{m} \frac{\Delta t}{2}$, and $\Delta t$ is the PIC time step.
Note that the calculation of $\vec{v}_{k+1}$ requires the value of the electric field at the time step $k+1$.
This is obtained in the field solver between steps \eqref{eq:leapfrog2} and \eqref{eq:leapfrog3}, using as input the positions $\vec{r}_{k+1}$.

Since we have used periodic BCs for the electric field in the lateral directions, we apply the same type of symmetry for the SPs.
This means that an SP that crosses the side boundaries $\Gamma_2$, will reappear at the symmetric point at the opposite boundary of the box.
Such a mapping, however, occurs extremely rarely, as the simulation box is chosen wide enough to prevent particles reaching sides before hitting the top boundary $\Gamma_1$ where they are absorbed.
For that reason, the contribution of mirror images of the simulation box to the total SP flux is negligible.
Note that the top boundary with a Neumann BC does not represent a physical anode electrode, but rather the uniform far field behavior of the gap between two macroscopically flat electrodes. 
Thus the box is chosen in our simulation to be sufficiently high so that the electric field becomes uniform and the space-charge negligible, which motivates the SP removal at that side.

Next to the top boundary $\Gamma_1$, the cathode boundary $\Gamma_3$ also acts as a particle absorber.
On $\Gamma_3$, the absorbed particles may cause secondary effects, such as secondary and higher order emission, plasma recycling, impurity sputtering, etc. Since these effects are beyond the scope of the current model and almost all electrons are removed at the top boundary, we leave their implementation for future work. Some attempts to take into account the above-mentioned effects can be found elsewhere \cite{brooks_sheath_2000,taccogna_effects_2004,taccogna_negative_2007,taccogna_plasma-surface_2004,taccogna_plasma_2005}.

\subsubsection{Particle weighting}
\label{sec:particle_weighting}
An important part of any PIC simulation, both conceptually and computationally, is mapping of the electric field to the position of SPs.
In our model, we first locate the finite element that surrounds the particle and then use the shape functions of that element to interpolate the electric field at the position of the SP.
The search for the element is carried out in a similar manner as during the particle injection (see appendix \ref{sec:rnd_injection}), but instead of the three barycentric coordinates, here we need four, i.e., one for each tetrahedral vertex.

To solve eq. \eqref{eq:poisson} by solving a matrix equation \eqref{eq:poisson_mat}, an inverse operation needs to be performed since the SP charge is required in the right-hand side (RHS) $f_i$ of eq.~\eqref{eq:poisson_mat}.
This, however, cannot be done immediately, as the SP in PIC is assigned a discrete charge $q$, while in eq. \eqref{eq:poisson} the charge appears in a form of charge density $\rho$.
To overcome this issue, we represent the charge of each SP by means of the $\delta$ function, so that
\begin{equation}
\label{eq:q_delta}
\rho(\vec{r}) = \sum_i q_i \delta(\vec{r}-\vec{r}_i) \comma
\end{equation}
where $\delta(\vec{r}-\vec{r}_i)$ is the Dirac $\delta$ function and the summation goes over the SPs that are located in the finite element where $\rho$ is evaluated. 
By doing such a substitution, the RHS of eq. \eqref{eq:poisson_mat} becomes
\begin{equation}
\label{eq:poisson_rhs}
f_i = \frac{q w_{sp}}{\epsilon} \sum_j N_i(\vec{r}_j) + \int_{\Gamma_1} N_i E_0 d\Gamma \comma
\end{equation}
where $\vec{r}_j$ are the coordinates of the $j$th SP that is located in the element where $f_i$ is computed.
It is important to note that to prevent self-forces, the same shape functions $N_i$ are used as during the mapping of the field to the position of the SPs and during the distribution of the charge inside an element.

\subsubsection{Particle injection}
\label{sec:particle_injection}
In our simulations, the space-charge is built up due to intensive electron emission.
Using the field emission tool GETELEC \cite{kyritsakis_general_2017}, we calculate the current density $J_e$ in the centroid of each quadrangle that is located at the metal-vacuum boundary.
$J_e$ determines the number of electron SPs, $n_{sp}$, that will be injected from a given quadrangle with area $A$ at given time step:
\begin{equation}
\label{eq:nsp}
n_{sp} = \frac{J_e A\Delta t}{ew_{sp}} \comma
\end{equation}
where $e$ is the elementary charge.
In general, $n_{sp}$ is a real number that has integer and fractional parts.
Since the number of injected SPs can be only integer, we use a uniformly distributed random number $R \in [0,1]$ to decide whether to round $n_{sp}$ up or down:
\begin{equation}
\label{eq:rnd_nsp}
\begin{aligned}
n_{sp}^{'} &= \text{floor}(n_{sp}) && \text{if } R \ge n_{sp}- \text{floor}(n_{sp})\\
n_{sp}^{'} &= \text{ceil}(n_{sp}) && \text{otherwise.}
\end{aligned}
\end{equation}

This discretization scheme guarantees that on average, the number of emitted SPs remains the same as given by eq.~\eqref{eq:nsp}.
The emitted SPs are distributed uniformly on the quadrangle surface. More details about it can be found in appendix \ref{sec:rnd_injection}.
After injection, the electric field $\vec{E}=\vec{E}(\vec{r})$ is used to give them an initial velocity
\begin{equation}
\label{eq:sp_velocity}
\vec{v}_0 = \frac{q}{m} \vec{E} \Delta t \left( \frac{1}{2} + R \right)
\end{equation}
and displacement
\begin{equation}
\label{eq:sp_displacement}
\Delta \vec{r} = \vec{v}_0 \Delta t R \period
\end{equation}

We add a term $\frac{1}{2}$ in eq.~\eqref{eq:sp_velocity} to accelerate the particle by half a time step and initialize the velocity for proper leapfrog stepping.
It is needed, because to enhance computational efficiency, we inject SPs between particle mover steps of \eqref{eq:leapfrog1} and \eqref{eq:leapfrog2} that were described in sec.~\ref{sec:particle_mover}.
The additional random number $R$ provides a \textit{fractional push} by a small random fraction of the previous time step, preventing the particles from forming artificial bunches \cite{timko_field_2015}.
This small initial push enhances computational stability, without significantly affecting the results, as shown in sec.~\ref{sec:validation}.

\subsubsection{Electron superparticle collisions}
\label{sec:sp_collisions}
In addition to interaction with the field, every electron experiences the effect of other charges present in the system.
A naive approach to take into account the Coulomb forces would be to calculate exactly the interaction between each pair of particles within the Debye sphere.
This, however, would result in $\mathcal{O}(n^2)$ complexity and is therefore not suitable for the large systems aimed in the present work.
Instead, we applied a more reasonable Monte Carlo binary collision model \cite{takizuka_binary_1977}. More details about it can be found from appendix \ref{sec:collisions}.

\subsection{Electrostatic forces}
The charge induced by an electric field on the metal surface affects the dynamics of the atomic system due to additional electostatic interactions.
To assess this effect, the value of the induced charge on the atoms must be estimated and introduced to the MD simulations where electrostatic interaction between the atoms is included as a force perturbation.
In this way the total force on an atom becomes $\vec{F} = \vec{F}_{EAM} + \vec{F}_L + \vec{F}_C$, where $\vec{F}_{EAM}$ is the force obtained from the interatomic potential.
The main perturbation is due to the Lorentz force
\begin{equation}
\label{eq:lorentz_force}
\vec{F}_L(\vec{r})=\frac{1}{2} q\vec{E}(\vec{r}) \period
\end{equation}
In this equation, the term~$\frac{1}{2}$ originates from the Maxwell stress tensor \cite{landau_electrodynamics_1984} and takes into account that only half of the charge is exposed to the field; the rest is shadowed by the material domain.
A less significant contribution to the total force $\vec{F}$ comes from the Coulomb interactions between the surface charges
\begin{equation}
\label{eq:coulomb_force}
\vec{F}_C(\vec{r}_i) = \frac{1}{4\pi\epsilon_0} \sum_{j\neq i} \frac{q_i q_j}{r_{ij}} \hat{r}_{ij} \exp(-\xi r_{ij}) \comma
\end{equation}
where $q_i$ is the charge of $i$th atom, $r_{ij}$ the distance between the $i$th and $j$th atoms, and $\hat{r}_{ij}$ the unit vector in the direction of $\vec{r}_{ij}$.
The exponent in eq.~\eqref{eq:coulomb_force} describes the screening of the $i$th charge by the conduction electrons in metals and the value of the screening parameter $\xi$ depends on the material and its crystallographic orientation. In our case, $\xi(\text{Cu})=0.6809$ \AA \supers{-1} as defined in \cite{djurabekova_atomistic_2011}.

The surface charge can be estimated according to Gauss's law \cite{djurabekova_atomistic_2011}
\begin{equation}
    \label{eq:gauss_law}
    \oint_{\Gamma} \lvec{E}\cdot \lvec{dA} = \frac{Q}{\epsilon_0} \comma
\end{equation}
where $Q$ is the charge inside a closed surface $\Gamma$.
Using eq.~\eqref{eq:gauss_law}, one can discretize the continuous distribution of surface charges on a metal surface to assign a partial charge to each atom on the surface corresponding to the value of the electric field applied at the atom position.
However, it is not trivial to decide how to identify the area which contributes to the charge on a given atom. 
In~\cite{djurabekova_atomistic_2011,kyritsakis_thermal_2018}, a rectangular geometry of the grid points associated with each atom was suggested.
Yet, this approach is limited by a high computational cost for large-scale simulations.
In the current work, we selected a different approach to calculate the surface charge, which allowed for higher computational efficiency while maintaining satisfactory accuracy.
This approach consists of two steps.
First, we calculate the charge of the mesh faces in the metal-to-vacuum boundary, hereafter referred to as face-charge:
\begin{equation}
    \label{eq:face_charge}
    Q_i = \epsilon_0 E_i A_i \text{, } i=1,2,..., n_{faces} \period
\end{equation}
Notice, that in eq. \eqref{eq:face_charge} it is enough to use the scalar field and area, since $\vec{E}\parallel \vec{A}$ due to the Dirichlet BC.
By using a weight function $w_{ij}\equiv w(\vec{r}_i-\vec{r}_j)$ (not to be confused with the PIC superparticle weight $w_{sp}$), the face-charge is distributed between the atoms as 
\begin{equation}
    \label{eq:atom_charge}
    q_j=\sum_i w_{ij} Q_i \text{, } j=1,2,..., n_{atoms} \period
\end{equation}
The total charge must remain conserved, therefore the weight function must satisfy
\begin{equation}
    \label{eq:charge_cons}
    \sum_i w_{ij} = 1 \text{ } \forall \text{ } j \period
\end{equation}
As the exact mathematical form of $w_{ij}$ can be chosen quite arbitrarily, we chose a form that is computationally cheap and prevents singularities for close points:
\begin{equation}
    \label{eq:charge_weight}
    w_{ij} = \frac{ \exp(-|\vec{r}_i-\vec{r}_j| / r_c) }
    { \sum_j  \exp(-|\vec{r}_i-\vec{r}_j| / r_c) } \comma
\end{equation}
where the cutoff distance $r_c$ determines the range where the charge $Q_i$ is distributed.
The tests show that on flat regions the results are almost independent of the exact value of $r_c$ until it exceeds the maximum edge length of surface triangles.
For this reason in our simulations we use an empirical $r_c$ value of 0.1 nm.

\begin{figure}[h!]
    \includegraphics[scale=0.75]{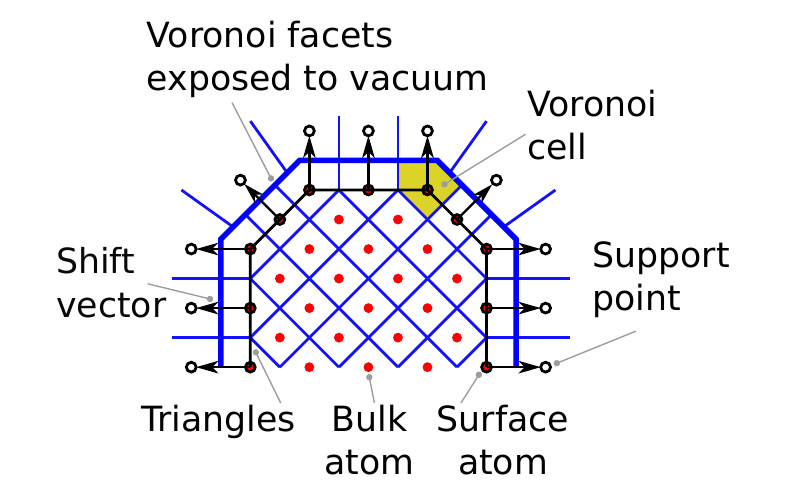}
    \caption{Generation and usage of Voronoi cells for surface charge calculation.}
    \label{fig:voronois}
\end{figure}

The face-charge method allows calculating an accurate surface charge for atoms that are located on atomic planes.
Charges in the regions with protruding features require further effort to achieve satisfactory accuracy.
We improve the accuracy by building the Voronoi tessellation around the surface atoms in those regions as shown in fig.~\ref{fig:voronois}.
The facets of the Voronoi cells are assumed to constitute the surface area for the given atom to estimate the charge on it.
Before generation, however, additional support points need to be built above the surface to limit the extension of the Voronoi cells into the infinity.
We build such support cloud by the following process:
\begin{enumerate}[noitemsep]
    \item Copy surface atom as a new point $\vec{p}$;
    \item Locate the triangle $f$ where the projection of $\vec{p}$, in direction of $f$ normal $\vec{n}$, lies within $f$;
    \item Move $\vec{p}$ in the direction of $\vec{n}$ for distance $\lambda$;
    \item Loop until each surface atom has a support point.
\end{enumerate}

Empirical tests have shown that good results are obtained if the length of the shift vector $\lambda$ equals one lattice constant of the atomistic system.
The coordinates of the surface atoms and the support points are input in the TetGen \cite{si_tetgen_2015} to generate a Voronoi tesselation around these atoms.
Note that the bulk atoms are in fig. \ref{fig:voronois} shown for visual purposes only, and are not considered during the Voronoi tesselation generation.

For each surface atom, we determine the Voronoi facets that are exposed to the vacuum, which are used in eq.~\eqref{eq:gauss_law} to assess the partial charge associated with the atom.
Assuming that the electric field in the location of the facets equals approximately the field at the atom position, $\vec{E}_i \approx \vec{E}_j$, we approximate eq. \eqref{eq:gauss_law} for atoms in the regions with increased roughness as
\begin{equation}
    \label{surf_charge}
    q_j = \epsilon_0 \vec{E}_j \cdot \sum_{i} \vec{A}_i \comma
\end{equation}
where the summation goes over the Voronoi facets of the $j$th atom that are exposed to vacuum.

\subsection{Heating}
We have shown previously \cite{parviainen_electronic_2011,eimre_application_2015,kyritsakis_thermal_2018} that thermal effects caused by field emission play a significant role in the evolution of nanotips under a high field.
To take these effects into account, we calculate the electron emission current density $J_e$ and the Nottingham heat $P_N$ \cite{charbonnier_nottingham_1964,paulini_thermo-field_1993} on the emitting surface by using our field emission tool GETELEC \cite{kyritsakis_general_2017}. 
To calculate the volumetric resistive heating power density, we use Joule's law as
\begin{equation}
\label{eq:joule}
P_J(T) = \sigma(T)(\nabla \Phi)^2 \comma
\end{equation}
where $\sigma(T)$ is the electric conductivity and $\Phi$ is the electric potential obtained by solving the continuity equation in a metal (domain $\Omega_2$ in fig.~\ref{fig:boundary})
\begin{equation}
\label{eq:cont}
\grad \cdot (\sigma \grad \Phi) = 0 \comma
\end{equation}
which comes with the following boundary conditions:
\begin{equation}
\label{eq:cont_bc}
\begin{aligned}
\sigma \grad \Phi = \vec{J_e} &\text{ on } \Gamma_3 \comma\\
\sigma \grad \Phi = 0 &\text{ on } \Gamma_4 \comma\\
\Phi = 0 &\text{ on } \Gamma_5 \period
\end{aligned}
\end{equation} 

The resistive and Nottingham heat cause a nonuniform temperature distribution inside the tip.
This distribution is found by solving the time-dependent heat equation in a metal (domain $\Omega_2$)
\begin{equation}
\label{eq:heat}
\grad \cdot (\kappa \grad T) + P_J = C_v \dt{T} \comma
\end{equation}
where $C_v$ is the volumetric heat capacity.
The initial and boundary conditions for eq. \eqref{eq:heat} are
\begin{equation}
\label{eq:heat_bc}
\begin{aligned}
T(t=t_0) = T_{amb} &\text{ on } \Omega_2 \comma\\
\kappa \grad T \cdot \vec{n} = P_N &\text{ on } \Gamma_3 \comma \\
\kappa \grad T \cdot \vec{n} = 0 &\text{ on } \Gamma_4 \comma\\
T = T_{amb} &\text{ on } \Gamma_5 \comma
\end{aligned}
\end{equation}
where $T_{amb}$ is the ambient temperature and $P_N$ is the boundary heat flux due to the Nottingham heating.
In eq. \eqref{eq:heat} and \eqref{eq:heat_bc}, $\kappa = \kappa (T)$ is the heat conductivity as given by the Wiedemann-Franz law \cite{franz_ueber_1853}
\begin{equation}
\label{eq:wied_franz}
\kappa(T) = LT \sigma(T) \comma
\end{equation}
where $L$ is the Lorentz number (see the next paragraph).
Equations \eqref{eq:cont} and \eqref{eq:heat} with the corresponding boundary conditions are solved in FEMOCS by means of FEM.
The details about how we discretized these equations can be found in appendices \ref{sec:weak_cont} and \ref{sec:weak_heat}.

Since we aim to simulate the dynamic evolution of nanotips, the mean free path of electrons inside such systems becomes comparable with the dimensions of the tip itself. 
Similarly to our previous work \cite{kyritsakis_thermal_2018}, we took into account the finite-size effects (FSEs) by modifying the bulk value of $\sigma(T)$ \cite{schuster_improved_2001} by a correction factor $\nu=\nu(T,d)$, as calculated by a simulation method developed by Yarimbiyik \textit{et al.} \cite{yarimbiyik_modeling_2006}.
The characteristic size of the nanotip $d$ equals the mean diameter along the tip for the initial geometry.
Similarly, for the Lorentz number in eq. \eqref{eq:wied_franz} we use an FSE-reduced value of $L = 2.0 \cdot 10^{-8}$  W $\Omega$ K\supers{-2}, as reported for a Cu film of 40 nm thickness \cite{nath_thermal_1974}.

The resulting temperature distribution is used to adjust the atomistic velocities in MD.
For this, the atoms are first grouped according to mesh tetrahedra that surround them.
After this, a Berendsen thermostat \cite{berendsen_molecular_1984} is applied separately to each such group.
The target temperature $T$ of the thermostat is obtained by averaging nodal temperatures of a tetrahedron, while the macroscopic temperature of the atomistic domain $T_0$ equals the average of microscopic temperatures of all the atoms of the same group.
For the control, we use a time constant of $\tau=1.5$~ps.
Such $\tau$ has shown previously \cite{kyritsakis_thermal_2018} to prevent the simulation artifacts in MD, while being significantly lower than the relaxation time of the heat equation.

\section{Results and discussion}
\label{sec:results}

\subsection{Model validation}
\label{sec:validation}
Prior to using the proposed model for actual simulations of physical systems, we validate our space-charge and surface charge calculation methods.
The field solver and emission module were verified earlier in \cite{veske_dynamic_2018} and \cite{kyritsakis_general_2017}.

\subsubsection{Validation of PIC}
We validate our PIC model by calculating the current-voltage curve for a planar cathode-anode system, for which a 1D semianalytical solution of Poisson's equation is available \cite{child_motion_1911,barbour_space-charge_1953,forbes_exact_2008}.
For this, we apply a voltage $V$ between two parallel plane electrodes of area $A=135$~nm$^2$ that are separated by a gap distance of $d=$18.2~nm and run PIC simulations for a few tens of fs until the total emission current $I$ converges to a steady-state value.

For this geometry, the current density $J$ can be accurately calculated by utilizing a stationary-point iteration to obtain the self-consistent values of the cathode electric field as obtained by the analytical solution of Poisson's equation and the current density $J$ as found by the Fowler-Nordheim equation \cite{kyritsakis_thermal_2018}. 
For high voltages, this curve asymptotically converges to the Child-Langmuir law \cite{child_motion_1911,langmuir_effect_1913,langmuir_effect_1923}.

\begin{figure}[h!]
    \includegraphics[scale=0.92]{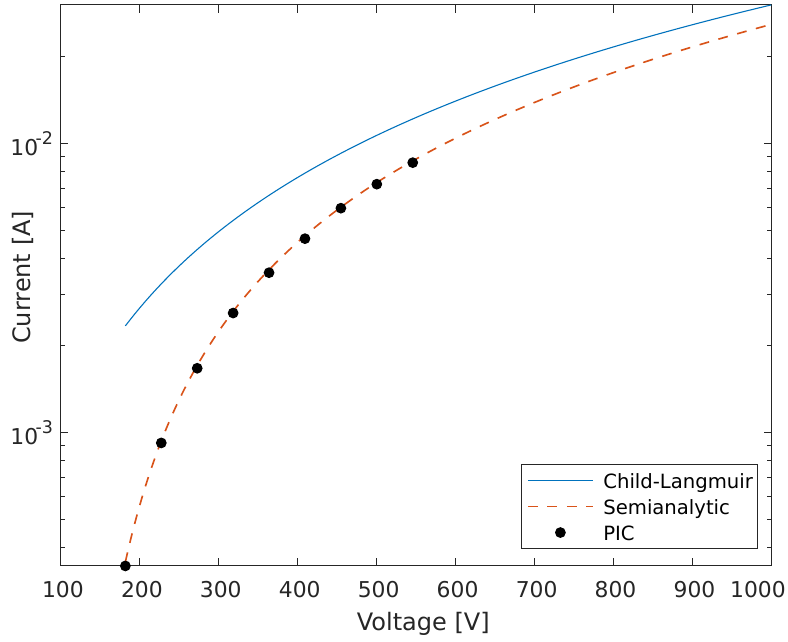}
    \caption{Current-voltage dependence from various models.}
    \label{fig:pic_validation}
\end{figure}

The results of the PIC simulation, together with the semianalytical and the Child-Langmuir curves, are shown in fig. \ref{fig:pic_validation}.
We observe that our PIC model is in very good agreement with the semianalytical curve with an RMS error of 1.9\%.

\subsubsection{Validation of surface charge model}
To validate the surface charge calculation method, we compare it against our earlier model HELMOD \cite{djurabekova_atomistic_2011}.
For this, we calculate the charge distribution on the apex of an $R=3$~nm, $h=11.5$~nm nanotip that is placed on a $W=93.2$~nm substrate.
Both the tip and substrate are cut out from a single-crystalline $\left<100\right>$ Cu block.

\begin{figure}[h!]
    \includegraphics[scale=1.0]{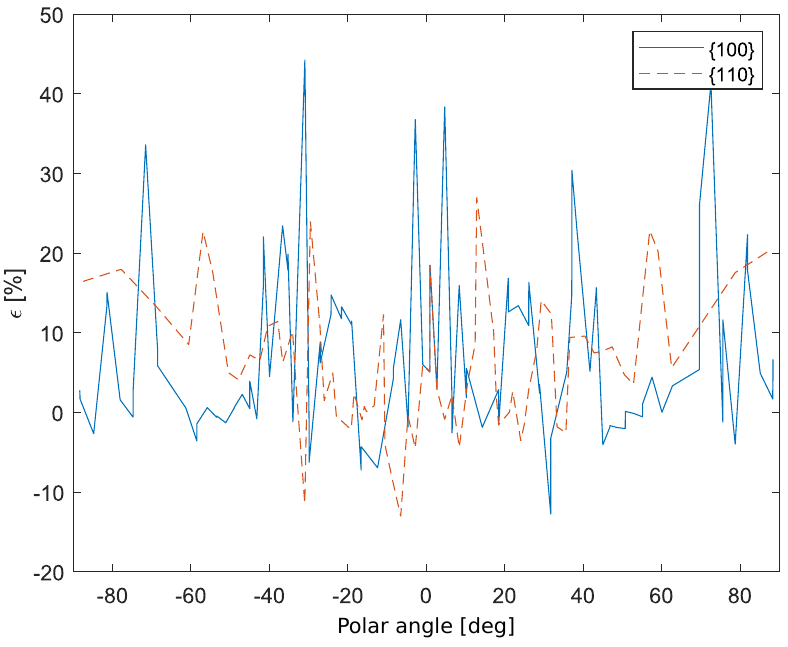}
    \caption{Relative difference between the surface charges obtained with HELMOD \cite{djurabekova_atomistic_2011} and with the current model. The difference is calculated along the surfaces of a $\{100\}$ and a $\{110\}$ slice of the apex.}
    \label{fig:charge_validation}
\end{figure}

We measure the surface charge both with HELMOD and with the model presented here and calculate the relative difference $\epsilon$ between them.
The variation of $\epsilon$ along the surface of the $\{100\}$ and $\{110\}$ slice of the apex is presented in fig. \ref{fig:charge_validation}.
The data show that the current model tends to provide slightly smaller charges than HELMOD with a mean difference of $6.5\pm1$\%.
The largest deviation occurs on sites that correspond to atoms below atomic steps and kinks.
This indicates that Voronoi cells that are built around those sites tend to be poorly exposed to vacuum, and for that reason, a smaller surface area is assigned for those sites as compared with HELMOD.
From the $\{110\}$ slice it is also visible that $\{110\}$ surfaces are a source of systematic difference.
This occurs because the Voronoi cell-based model tends to assign charge also for atoms on the second layer of the $\{110\}$ surface, while HELMOD considers those atoms as totally shadowed.

\subsection{Simulation setup}
\label{sec:simulation_setup}
We used the model described in sec. \ref{sec:method} to simulate the behavior of a Cu nanotip under a high electric field.
The expected outcome of the simulation is that the upper part of the tip will experience extensive deformations, while the lower section remains practically constant.
Such a scenario helps us to optimize the computational cost by simulating in MD only the upper half of the initial system.
The bottom region, which completes the geometry for calculating the physical quantities that affect the atom motion, is built separately and remains unchanged for the whole simulation.
We choose different widths for this extension in order to investigate the impact of the heat dissipation on the processes at the apex of the tip.

The geometry of the tip that defines the vacuum domain~$\Omega_1$ and the metal domain~$\Omega_2$ is illustrated in fig.~\ref{fig:boundary}.
In~$\Omega_1$, PIC and electric field calculations are performed, while MD and the heat solver operate in $\Omega_2$.
Notice, that MD uses only the upper part of $\Omega_2$ (enclosed with dashed lines in fig. \ref{fig:boundary}), while FEM covers all of it.
A~conical tip with an aperture angle $\gamma=3^\circ$, initial height $h=93$~nm and a hemispherical cap of radius $R=3$~nm is placed into a simulation box of height $H=10h$ ($H$~is adjusted with the change in nanotip height).
The tip lies on a substrate of width $W=620$~nm and height $H_b=7.3$~nm.
The atomistic region is cut out from a monocrystalline $\left<100\right>$ Cu box with a lattice constant of 3.64~\AA.
The bottom layer of the atoms is fixed in place to ensure a smooth transition to the extended part.
Two different extension geometries were used, with characteristic radii of $r_{thin}=17$~nm and $r_{wide}=54$~nm.

The simulation geometry is motivated from the experimental observations that suggest the existence of field emitters with an aspect ratio of 20-100 \cite{kildemo_new_2004} on flat Cu surfaces.
The size of the tip is chosen to be large enough to prevent instability due to excess surface energy  \cite{veske_electrodynamics-molecular_2016,jansson_long-term_2016} and sufficiently small to result in a reasonable computational cost.
In order to obtain results that are directly comparable to our previous work \cite{kyritsakis_thermal_2018}, we chose the dimensions of the thin system to be identical to the ones used in \cite{kyritsakis_thermal_2018}.
In order to obtain sufficient mesh density, the characteristic distance between mesh nodes in the apex region is chosen to be one half of lattice constant of the initial atomistic system.

The MD simulations were carried out by means of the classical MD code PARCAS \cite{nordlund_molecular_1994,nordlund_defect_1998,ghaly_molecular_1999}.
The FEM calculations in FEMOCS are based on open-source C++ library Deal.II \cite{bangerth_deal.ii_2007}.
For the MD simulations we used the interatomic EAM potential developed by Mishin \textit{et al.} \cite{mishin_structural_2001}.
This potential has been successfully used in our previous works \cite{kyritsakis_thermal_2018,veske_electrodynamics-molecular_2016,parviainen_electronic_2011} and has shown accurate reproduction of nonequilibrium system energetics \cite{mishin_structural_2001}.
The stochastic nature of the thermal effects were taken into account by running 50 independent simulations with different random seeds.
In all cases, the initial velocities were sampled from the Maxwellian distribution with $T=300$ K.
No periodic boundaries were applied for the MD cell, while in FEM and PIC calculations, the periodic boundaries were applied in lateral dimensions.
A constant time step of 4.05 fs was used for MD, 0.51 fs for PIC and 40 fs for the heat equation.
An electron SP weight of $w_{sp}=0.01$ in PIC gave a sufficiently dense sampling of the phase space, in order to ensure a smooth space-charge distribution.

We applied a long-range electric field $E_0=0.6$ GV/m for the thin tip.
The lower field enhancement factor of the wide tip as compared to the thin one is compensated by adjusting the applied field to $E_0=0.608$ GV/m to ensure the same field emission current from both tips.

\subsection{Course of the simulation}

We simulated the runaway process in a nanotip similarly to the one in \cite{kyritsakis_thermal_2018}.
Despite the usage of a more simplistic 1D space-charge model in \cite{kyritsakis_thermal_2018}, the overall picture of the runaway process was confirmed in current simulations.
In addition, here we provide the statistical analysis of the nanotip behavior and estimate the duration of the entire process until the tip stops emitting neutrals.

\begin{figure*}
    \includegraphics{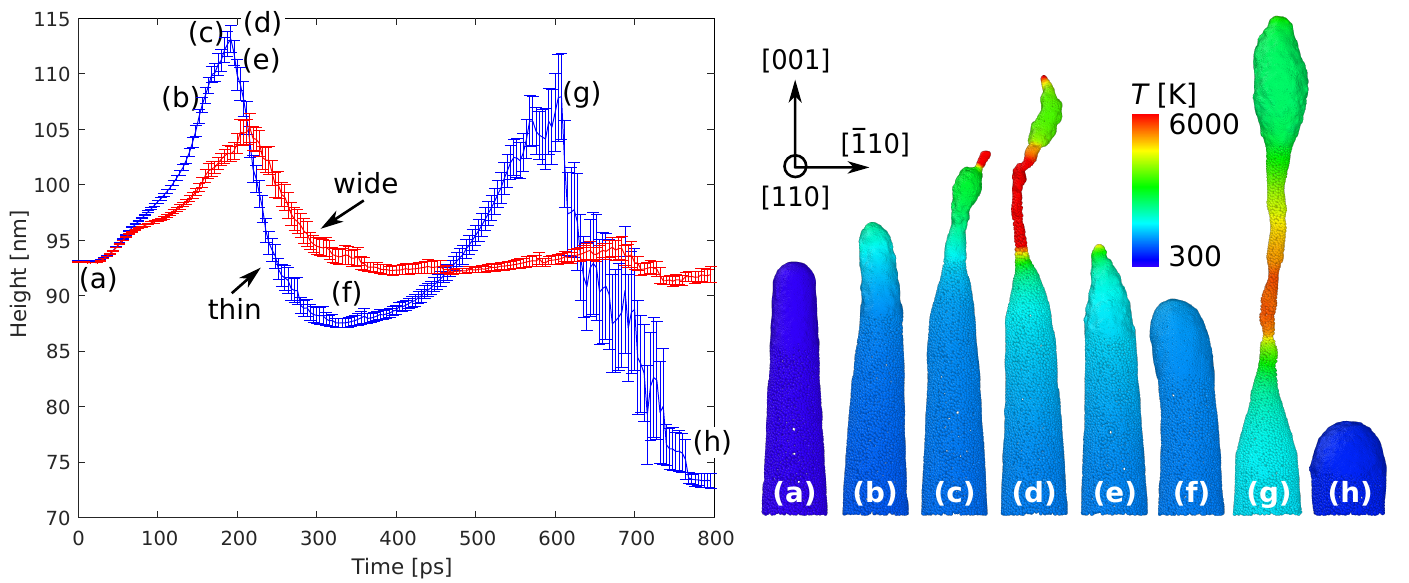}
    \caption{Time evolution of the averaged nanotip heights (left) and some characteristic excerpts from the simulation (right).
        Labels (a)-(h) on the left correspond to the frames on the right side.
        Error bars show the variation of the data between parallel runs in a form of standard error of the mean.}
    \label{fig:height}
\end{figure*}

In fig. \ref{fig:height}, we show the evolution of the height of the thin and wide nanotips that is averaged over the parallel runs.
The error bars in this graph indicate the standard error of the mean height and reflect the differences in evolution of individual tips.
On the graph one can observe a tendency in height variations over time in the form of a wave.
Due to the high temperatures and strong forces the droplets of molten Cu once in a while leave the apex of the tip.
Normally this leads to an abrupt change in the tip height, but due to averaging this change is smoothed producing the shape of a wave.
A rise of the tip height after its fall is seen in the majority of the simulations, indicating that the runaway is able to revive.
This tendency is stronger for thinner tips, since the efficient heat conduction in the wider tips cools the shortened tip faster.
To illustrate the process of height change in the tips evolving under the applied electric field, some characteristic simulation snapshots designated by the corresponding letters on the left of fig. \ref{fig:height}, are provided on the right of fig. \ref{fig:height}.

\begin{figure*}
    \includegraphics{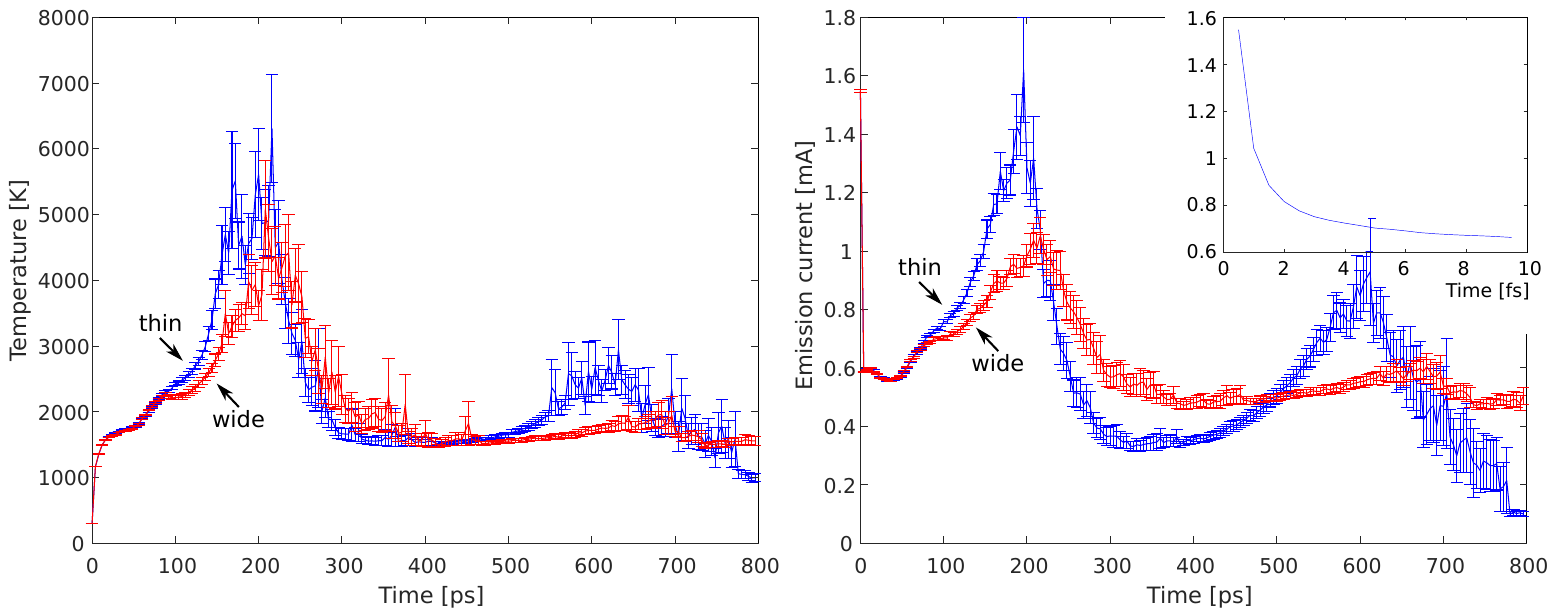}
    \caption{Time evolution of averaged apex temperature (left) and total field emission current (right) from the nanotips. The inset shows the magnification of the first 10 fs, during which the space-charge is formed.}
    \label{fig:temp_current}
\end{figure*}

In fig. \ref{fig:temp_current}, we analyze the correlation between the averaged total emitted current and the apex temperature.
In this graph, we see that due to the build-up of space-charge, the field emission current drops suddenly within the first few fs of the simulation.
In the following approximately 20-25 ps, there is another slight but consistent drop in the current.
This feature appears due to the shape modification of the apex region after melting, which leads to a reduction of the emitting area.

As long as the temperature in the tips does not exceed the melting point, the height of the tips does not change (fig. \ref{fig:height}a).
After the melting point was reached, the tips start growing, causing a step-by-step increase in the local field, current, and temperature (fig. \ref{fig:height}b).
The increase lasts until the thermal energy of the apex atoms exceeds their binding energy, which causes a gradual disintegration of the tip, either atom by atom, or in a form of clusters.
The latter is often preceded by the formation of a neck below the apex (fig. \ref{fig:height}d).
In some cases, a smaller cluster of the size up to a few atoms might also be detached (fig. \ref{fig:height}c).
Within a necking region, the high current density causes intensive Joule heating leading to a self-amplifying increase in local temperature until the top of the tip is detached.
After the break-up, the tips cool down (fig. \ref{fig:height}e), but within a few tens of ps, the temperature increases again and the whole process reappears.
The cycle lasts until the enhanced local field is not sufficiently high to generate the field emission current capable to heat the tip above the melting point.
In this case, the tips stabilize and the runaway process ceases (fig. \ref{fig:height}h).

\begin{figure}[h!]
    \includegraphics[scale=1]{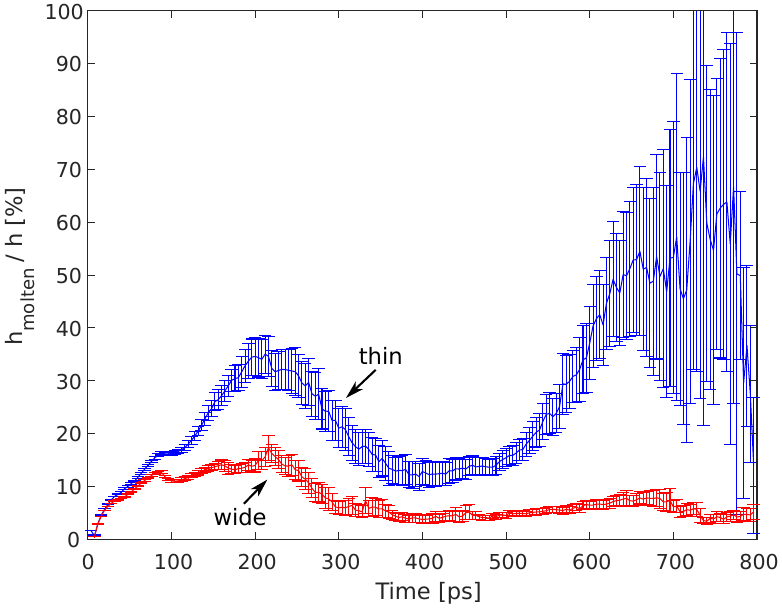}
    \caption{Time evolution of the relative height of molten region.}
    \label{fig:molten_height}
\end{figure}

Although initially the growth rate is similar in thin and wide tips, the temperature rise in the wide tip soon starts lagging behind the one in the thin tip.
The lag occurs despite the same amount of heat generated initially in both systems.
The more efficient cooling in the wider tip prevents the fast expansion of the region of molten Cu at the top of the tip, hence reducing also the growth rate.
This is seen in fig. \ref{fig:molten_height}, which shows the relative height of the molten region in a tip.
In this figure we see that in the thin tips, the intensive evaporation starts when about 30\% in the height of the tip is molten, remaining higher than this value for the whole evaporation period.
In the wide tip, on the other hand, that ratio barely reaches 15\%.
The molten region between solid Cu and the apex effectively acts as a thermo-electric insulator, since beyond the melting point, the thermo-electrical resistivity of Cu increases abruptly more than 60\% \cite{schuster_improved_2001}.
The higher percentage of the molten region in the thin tip reduces efficiently its heat conduction, inducing a stronger accumulation of the heat at the apex of the tip.

We note that figs. \ref{fig:height}-\ref{fig:molten_height} show the averaged data of multiple runs.
Due to averaging, we do not observe in these plots any abrupt transitions.
In these graphs, we observe a smooth modulation of the data.
We will refer to the well-pronounced peaks in fig. \ref{fig:height} as tip growth phases.
During each such growth phase the tip first sharpens, and after an evaporation event or a sequence of them, it blunts down until the emission currents re-heat the tip, restarting the growth phase.

\begin{figure}[h!]
    \centering
    \includegraphics[scale=1]{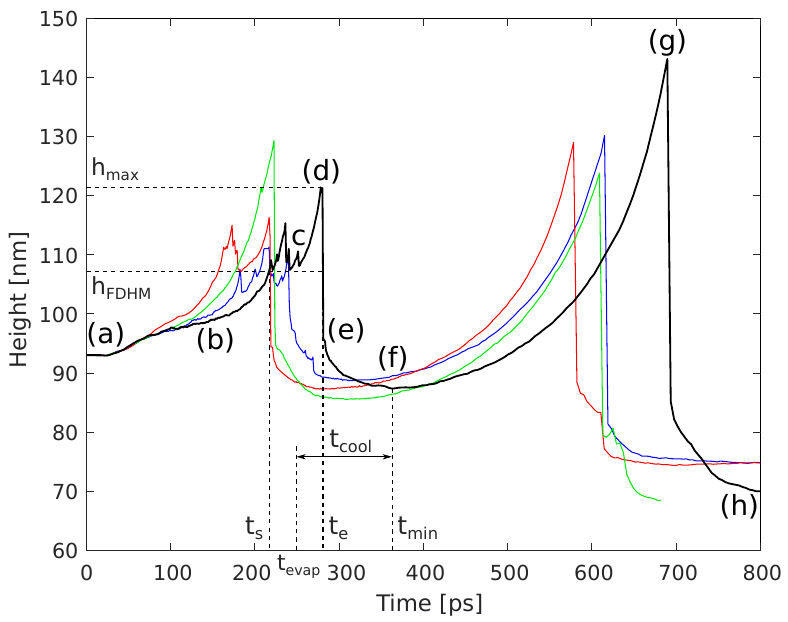}
    \caption{Four separate height profiles of thin nanotips that illustrate the variation in shape and timings of the curves. Dashed lines demonstrate the method for determining $t_{evap}$ and $t_{cool}$. Labels (a)-(h) correspond to snapshots in fig. \ref{fig:height}.}
    \label{fig:separate_heights}
\end{figure}

However, by plotting the height evolution of individual tips, we see that the height does not change smoothly, but rather abruptly due to large evaporation events, as demonstrated in fig. \ref{fig:separate_heights}.
Here, we see that the first growth phase does not have a single well-defined peak.
Instead, while the tip is still in the growing phase, smaller peaks indicating detachment of small clusters of atoms appear, although these do not affect dramatically the dynamics of the tip evolution.
These events are stochastic, and the time interval between them may vary from a few up to hundreds of ps.
Since evaporation starts already from the beginning of the growth phase (approximately when the first small peak occurs) and continues until the large piece of molten copper detaches from the surface, the time interval $t_{evap}$ until the major evaporation event is not well defined.
To describe quantitatively the dynamics of the field-emitting tip and enable the comparison between different cases, we measured the time intervals until the start and the end of the nanotip growth phase, $t_s$ and $t_e$, and defined $t_{evap}=\frac{1}{2}(t_s+t_e)$.
The time intervals $t_s$ and $t_e$ are found at boundaries of half maximum of the height curve around the first intensive evaporation period (see fig. \ref{fig:separate_heights}).
During the second tip growth phase the tip was growing until a large droplet of the molten copper was detached.
Since this was always a single event, we quantified the time interval until the second major evaporation event by determining the time when the tip reached its second maximum in its height.
We also analyzed the cooling time by determining the interval between the time instances when the tips reached their maximum and minimum heights and calculated $t_{cool}=t_{min}-t_{evap}$.
The results of these measurements are summarized in table~\ref{tab:data}.

As already mentioned, the two tip growth phases observed in thin tips resulted in evaporation processes with different characteristic features.
The first growth phase was accompanied by several evaporation events before the detachment of the large droplet, while during the second growth phase only one large disintegration occurred.
During the first growth phase, the heat is rapidly accumulating at the apex of the tip, increasing the kinetic energy of the atoms.
The pulling effect of the strong local electric field increases the probability of the energetic surface atoms to fly off the surface as single atoms or as small clusters of atoms.
After a few hundreds of ps, the temperature within the tip reaches an equilibrium, with a large fraction of the tip beneath the very top being molten.
Such a molten tip is flexible and responds gradually as a whole to the tensile stress exerted by the field.
This leads to stretching of the tip into the vacuum and a consequent necking.
The initial necking leads to the increase of the current density in the narrower regions, increasing the Joule heating and thus the local temperature in this region.
Eventually, a droplet forming above the neck flies off the surface leading to a large evaporation event, which completes the first growth phase.

During the second growth phase the apex temperature does not reach the same value as during the first one; see fig.~\ref{fig:temp_current}.
For that reason, during the second growth phase the tip grows gradually, and no small modulations of the height similar to the first growth phase are observed.
However, a larger fraction of the tip is already molten since the beginning of the second phase, rendering it flexible and responsive to the tensile stress exerted by the field.
This results in a necking under the top of the tip, forming a droplet since the beginning of the growth phase.
Hence, we observe only a single large evaporation event, which abruptly reduces the height of the tip.

It has been previously shown \cite{timko_field_2015} that a certain evaporation rate of neutral atoms is required in order to build up the vapor density necessary to trigger an ionization avalanche, which eventually leads to plasma formation.
In \cite{timko_field_2015}, the neutral evaporation rate $r_{Cu}$ was considered proportional to the electron emission rate $r_e$, with a minimum ratio of $r_{Cu/e}\equiv r_{Cu}/r_e$ of 0.015 required to lead to plasma formation. 
In our previous work \cite{kyritsakis_thermal_2018}, this ratio was found to have an average value of 0.025$\pm$0.003 over the whole evaporation process. 
Here we shall examine the evolution of the evaporation rate in greater detail.

\begin{figure}[h!]
\centering
\includegraphics[scale=1]{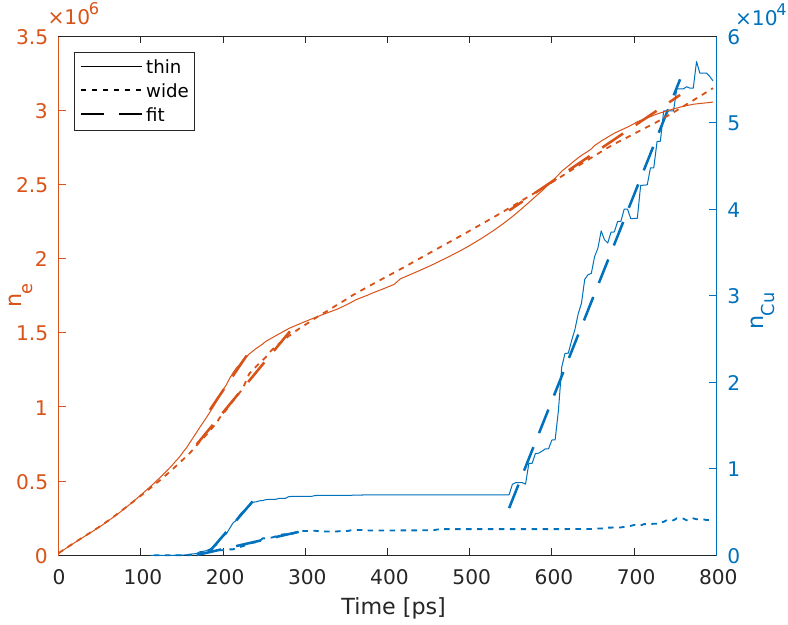}
\caption{Cumulative number of emitted atoms (right lines on the graph) and electrons (left lines). The slope of the linear fit (bold dashed lines) gives evaporation and injection rates.}
\label{fig:evap_inj_rate}
\end{figure}

For this purpose, we plotted in fig.~\ref{fig:evap_inj_rate} the average cumulative number of evaporated atoms and emitted electrons.
We observe that the runaway process consists of two alternating regimes, characterized by intensive evaporation and a metastable stage with no or very few evaporation events.
In both the thin and wide tips, the violent evaporation at constant rate remains to the intensive evaporation regime.
During this regime we observe large variations in height, current, field, and temperature.
In the metastable regime, on the other hand, the evaporation is practically not observed, while field emission continues.
In order to quantify the mean evaporation rate within the limits of each evaporation regime, we fit a straight line to each curve (bold dashed lines in fig.~\ref{fig:evap_inj_rate}), the slope of which gives the evaporation-emission rate.
The calculated parameters are summarized in table~\ref{tab:data}.

The data show, that thin tips start evaporating several tens of ps earlier than wide ones, but there is no significant difference in their cooling periods.
During the first growth phase, the thin tip is able to provide neutral atoms at a rate four to five times higher than the wide one, while the total currents differ only by 10-30\%.
From these results, we deduce that thin field emitters have higher probability to trigger plasma build-up than the wide tips.

\begin{table}[h!]
\centering
\caption{Results of the measurements.}
\label{tab:data}
\begin{tabular}{r|D{,}{\pm}{-1}D{,}{\pm}{-1}D{,}{\pm}{-1}c}
    & \multicolumn{1}{c}{wide} & \multicolumn{1}{c}{thin 1} & \multicolumn{1}{c}{thin 2} & \multicolumn{1}{c}{ArcPIC \cite{timko_field_2015}}\\
    \hline \\  [-0.8em]
    $t_{evap}$ [ps]            & 229,7       & 194,4     & 613,10   & \na \\
    $t_{cool}$ [ps]            & 176,18      & 175,20    & -        & \na \\
    $r_{Cu}$ [ps$^{-1}]$       & 22,2        & 110,5     & 238,12   & 404 \\
    $r_{e}$ [fs$^{-1}]$        & 6.7,0.2     & 8.2,0.6   & 3.7,0.2  & 27.1 \\
    $R_{Cu}$ [nm$^{-2}$ ns$^{-1}]$ & 220,20  & 1100,50   & 2380,120 & 40.4 \\
    $R_{e}$ [nm$^{-2}$ ps$^{-1}]$  & 67,2    & 82,6   	 & 37,2 	& 2.7 \\
    $r_{Cu/e}$ $[\times 10^{-3}]$  & 3.2,0.3 & 13.5,1.6  & 64,6     & 15 \\
\end{tabular}
\end{table}

\subsection{Discussion}
The implemented PIC module within the hybrid MD-FEM model allowed us to achieve a more accurate description of the emission currents, as compared to that given by the previously used simplified 1D space-charge model.
The model resulted in an increased emission for the same geometry and applied field, enabling us to observe the thermal runaway process at lower applied electric field $E_0 \approx 0.6$ GV/m, compared to 0.8 GV/m according to earlier estimations. 
This result brings the theoretical estimations closer to the experimental values, where vacuum breakdowns are commonly observed at external electric fields $E_0\leq 0.2$ GV/m \cite{descoeudres_investigation_2009}.

The observed reduction in the field and the growth in the current values are natural.
The simplified model is based on the equivalent planar diode approximation \mbox{\cite{barbour_space-charge_1953,Uimanov2011}}, which assumes a uniform current density in space.
While simulating field emitters, however, the emission current in space is highly nonuniform due to the nonplanar geometry, which is reducing the suppression of the local field due to the space-charge.

As stated earlier, we selected the applied field in our simulations, so that the total emission current at the beginning of the simulation in both thin and wide tips is the same.
We took special care for this to ensure that both tips develop initially under similar conditions.
However, we see that the evaporation rates from the thin and wide tips can differ by an order of magnitude.
Also, in the thin tips, we observed two growth phases, which resulted in two runaway events.
As it can be seen in table \ref{tab:data}, the ratio $r_{Cu/e}$ for the second runaway event is significantly higher than the threshold value of 0.015 that was determined in \cite{timko_field_2015}.
During the first runaway process the ratio $r_{Cu/e}$ is quite close to this value.
If we consider the average rates since the first evaporation event until the end of the simulation, we obtain a value $\left< r_{Cu/e} \right>=0.026\pm0.008$, which clearly exceeds the lower limit for triggering an avalanche ionization process.

On the other hand, for the wide tip, the ratio $r_{Cu/e}$ is significantly lower than the one assumed in \cite{timko_field_2015}. 
However, it is premature to conclude that thick field emitters are not able to produce enough neutral atoms to ignite plasma.
First, in the simulations performed by using the ArcPIC model in \cite{timko_field_2015}, a linear relationship between atom evaporation and electron emission rates was assumed.
Such a simplified assumption of the behavior of surface atoms in the pre-breakdown condition may affect the estimated threshold values severely.
It is evident from our results that the linear relationship between atom evaporation and electron injection rate is valid only as a rough qualitative approximation and the actual dependency is much more complex, with intermittent high evaporation events and metastable periods, for which electron emission continues.

Furthermore, in \cite{timko_field_2015}, a much wider electron emission and neutral injection area ($10^4$~nm$^2$) was assumed.
In the present simulations, we can safely consider that all emission and evaporation originate from an area smaller than 100~nm$^2$ (a hemisphere of 4~nm radius).
Given such a significant mismatch of the space scales, it is more reasonable to compare the surface evaporation and emission fluxes $R_{Cu}$ and $R_e$, i.e., the rates divided by the emission areas.
This is because it is the vapor and electron densities that determine whether the avalanche ionization reaction, needed to ignite plasma, can take place. 
These densities are not dependent on either the absolute overall evaporation or emission rates or their ratio.
They are rather connected more to the fluxes $R_{Cu}$ and $R_e$, the increase of which promotes plasma ignition.
From table~\ref{tab:data} we see that the values for $R_{Cu}$ and $R_e$ obtained from the present results significantly exceed the ones assumed in \cite{timko_field_2015}, meaning that even wide tips might provide an electron and neutral density sufficient to ignite plasma.

Although the runaway process and its connection to larger scale plasma onset simulations via the approximation of the $r_{Cu/e}$ coefficient or the evaporation flux $R_{Cu}$ offers a plausible explanation of the vacuum arc initiation, the exact mechanisms that lead to plasma formation are not yet clear.
Our current PIC model includes only electron SPs and omits the interaction of the evaporated material with the electrons and the appearance of positive ions that reduce the space-charge in the vicinity of cathode, increase field emission current, and trigger surface sputtering and heating.
However, the present developments form the basis for future expansion of the methodology to include more particle species in the PIC domain and properly handle the plasma-metal interactions.
This aspires to shed light to the detailed processes that lead to vacuum arc ignition, without the need for the rough qualitative approximations applied in \cite{timko_field_2015}.

\section{Conclusions}
We have investigated the dynamic evolution of Cu field emitters of different widths during the intensive field emission in the pre-breakdown condition.
For the study we developed a concurrent multiscale-multiphysics model that combines classical molecular dynamics, finite element method and particle-in-cell techniques.
The high efficiency and robustness of the model allowed us to perform extensive statistical analysis of a highly stochastic thermal runaway process.
The simulations show that the thermal runaway can start in a $h=93$ nm Cu nanotip under macroscopic electric fields of $E_0=0.6$ GV/m.
This value exceeds only three times the value repeatedly reported in experiments with flat copper electrodes.
In a thin field emitter, the thermal runaway is a cyclic process of alternating regimes of intensive atom evaporation and a subsequent cooling process of a tip.
The amount of evaporated atoms that are emitted from thin emitters is sufficient to ignite self-sustainable plasma.
Increasing the width of the emitter lowers the atom evaporation rate and decreases the probability for the occurrence of more than one runaway cycles.
Wide emitters show also a lower neutral evaporation rate, which leads to the conclusion that very sharp field emitters may be necessary to enable plasma formation.

\section*{Acknowledgements}
The current study was supported by the Academy of Finland project AMELIS (Grant No. 1269696), CERN CLIC K-contract (No. 47207461), Estonian Research Council Grant PUT 1372, and the Estonian national scholarship program Kristjan Jaak (No. 16-4.2/653), which is funded and managed by the Archimedes Foundation in collaboration with the Ministry of Education and Research of Estonia.
We also acknowledge grants of computer capacity from the Finnish Grid and Cloud Infrastructure (persistent identifier urn:nbn:fi:research-infras-2016072533).

\vfill
\pagebreak

\appendix

\section{Simulation flowchart}
\label{sec:femocs_long}

\vspace{-10pt}
\begin{figure}[h!]
  %  \centering
    \includegraphics[width=1.0\linewidth]{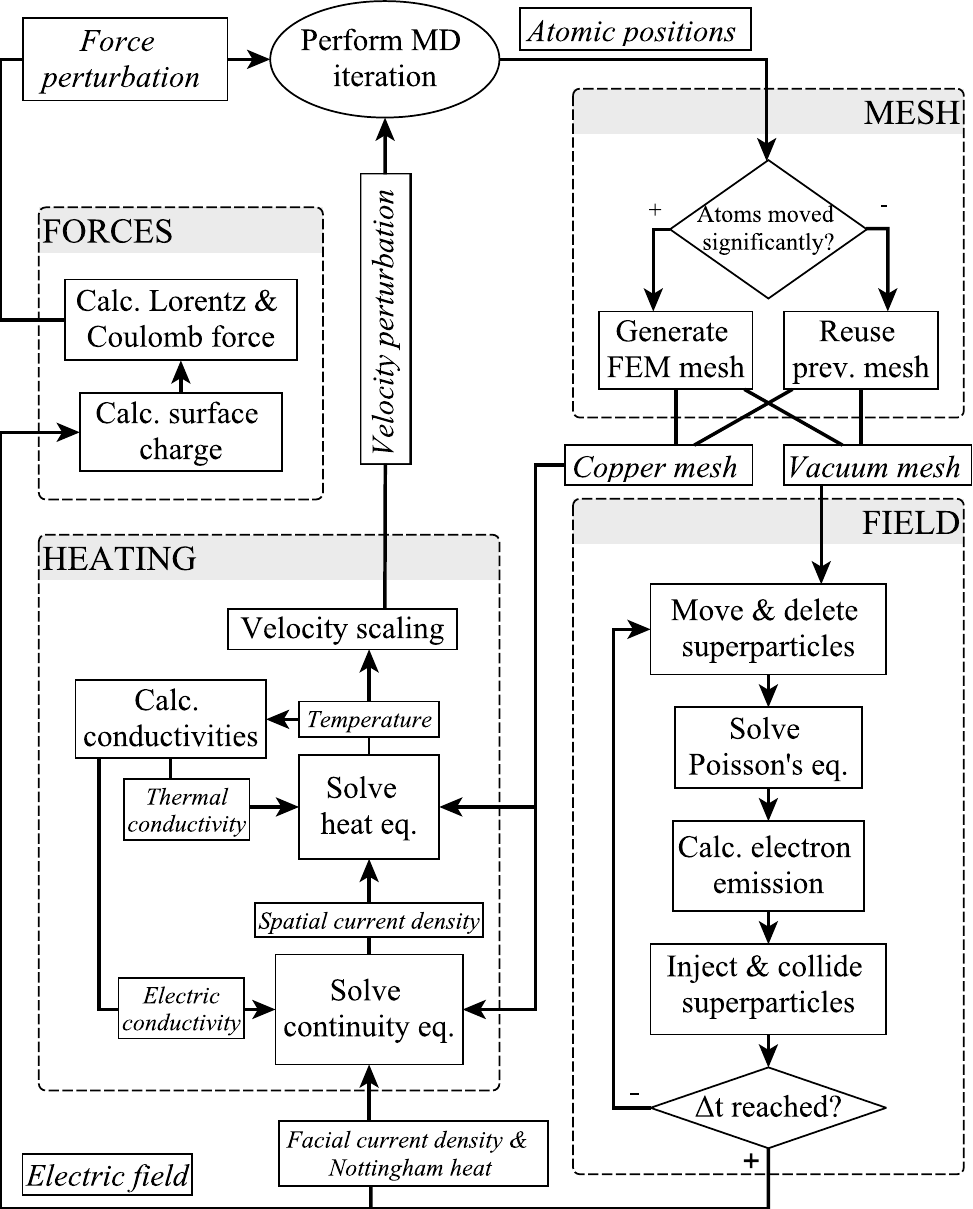}
    %\caption{Full schematics of multiscale-multiphysics model.}
    %\label{fig:femocs_long}
    %\vspace{-10pt}
\end{figure}

\section{Discretization of Poisson's equation}
\label{sec:weak_poisson}
Multiplying eq. \eqref{eq:poisson} with a weight function $w$ and integrating over the whole domain gives
\begin{equation}
    \label{eq:poisson_weak1}
    \int_{\Omega} w \grad \cdot (\epsilon \grad \Phi) d\Omega
    + \int_{\Omega} w \rho d\Omega = 0 \period
\end{equation}

Using the properties of differentials, the previous expression could be rewritten as
\begin{equation}
    \label{eq:poisson_weak2}
    \int_{\Omega} \grad \cdot (w \epsilon \grad \Phi) d\Omega
    - \int_{\Omega} \epsilon \grad w \cdot \grad \Phi d\Omega
    + \int_{\Omega} w \rho d\Omega = 0 \period
\end{equation}

The usage of Green’s theorem on the first term removes double differentiation of $\Phi$:
\begin{equation}
    \label{eq:poisson_weak3}
    \oint_{\Gamma} w \epsilon \grad \Phi \cdot \vec{n} d\Gamma
    - \int_{\Omega} \epsilon \grad w \cdot \grad \Phi d\Omega
    + \int_{\Omega} w \rho d\Omega = 0 \period
\end{equation}

By plugging BCs \eqref{eq:poisson_bc} into leftmost term, we obtain the weak form of eq. \eqref{eq:poisson}:
\begin{equation}
    \label{eq:poisson_weak}
    \int_{\Gamma_1} w \epsilon E_0 d\Gamma
    - \int_{\Omega_1} \epsilon \grad w \cdot \grad \Phi d\Omega
    + \int_{\Omega_1} w \rho d\Omega = 0 \period
\end{equation}

To transfer that formula into FEM format, we use the Galerkin approach and equalize the weight function with shape functions and discretize the potential as a linear combination of them:
\begin{equation}
    \label{eq:poisson_galerkin}
    \begin{aligned}
    w &= N_i, \quad i=1,2,..., n, \\
    \Phi(x,y,z;t) &= \sum_{j=1}^{n}N_j(x,y,z)\cdot \Phi_j(t) \comma
    \end{aligned}
\end{equation}
where $n$ is the number of degrees of freedom of the finite element.
By also combining $\epsilon$ and $\rho$ together, \eqref{eq:poisson_weak} becomes
\begin{equation}
\begin{split}
    \label{eq:poisson_mat1}
    \int_{\Omega_1} \sum_{j=1}^{n}(\grad N_i \cdot \grad N_j \Phi_j) d\Omega
    \\= \int_{\Omega_1} N_i \frac{\rho}{\epsilon} d\Omega
    + \int_{\Gamma_1} N_i E_0 d\Gamma \period
\end{split}
\end{equation}

To solve that system of equations numerically, it is handy to express it in matrix form
\begin{equation}
    \label{eq:poisson_mat}
    \vecb{M} \cdot \vecb{\Phi} = \vecb{f} \comma
\end{equation}
where
\begin{equation*}
    \begin{aligned}
    M_{ij} &= \int_{\Omega_1} \grad N_i \cdot \grad N_j d\Omega \comma\\
    f_i &= \int_{\Omega_1} N_i \frac{\rho}{\epsilon} d\Omega + \int_{\Gamma_1} N_i E_0 d\Gamma \period
    \end{aligned}
\end{equation*}

\section{Discretization of continuity equation}
\label{sec:weak_cont}
Notice the similarity between eq. \eqref{eq:cont} and \eqref{eq:poisson}.
Therefore, by replacing dielectric constant $\epsilon$ with conductivity $\sigma$ and by taking $\rho = 0$, we obtain the representative form of eq. \eqref{eq:poisson_weak3} for \eqref{eq:cont}:
\begin{equation}
    \label{eq:cont_weak1}
    - \int_{\Omega} \sigma \grad w \cdot \grad \Phi d\Omega
    + \oint_{\Gamma} w \sigma \grad \Phi \cdot \vec{n} d\Gamma = 0 \period
\end{equation}

The application of BCs \eqref{eq:cont_bc} into the second term gives the weak form of eq. \eqref{eq:cont}:
\begin{equation}
    \label{eq:cont_weak}
    - \int_{\Omega_2} \sigma \grad w \cdot \grad \Phi d\Omega
    + \int_{\Gamma_3} w \vec{J_e} \cdot \vec{n} d\Gamma = 0 \period
\end{equation}

To transfer that formula into algebraic form, we use the same strategy as in appendix \ref{sec:weak_poisson} and equalize weight function with shape functions and expand the potential as a linear combination of them.
In that way the eq. \eqref{eq:cont_weak} becomes
\begin{equation}
    \label{eq:cont_mat1}
    \int_{\Omega_2} \sigma \sum_{j=1}^{n}(\grad N_i \cdot \grad N_j \Phi_j) d\Omega =
    \int_{\Gamma_3} N_i \vec{J_e} \cdot \vec{n} d\Gamma \comma
\end{equation}
which looks in matrix form as
\begin{equation}
    \label{eq:cont_mat}
    \vecb{M} \cdot \vecb{\Phi} = \vecb{f} \comma
\end{equation}
where
\begin{equation*}
    \begin{aligned}
    M_{ij} &= \int_{\Omega_2} \sigma \grad N_i \cdot \grad N_j d\Omega \comma\\
    f_i &= \int_{\Gamma_3} N_i \vec{J_e} \cdot \vec{n} d\Gamma \period
    \end{aligned}
\end{equation*}

\section{Discretization of heat equation}
\label{sec:weak_heat}
To discretize the heat equation \eqref{eq:heat}, we use similar strategy as in appendix \ref{sec:weak_poisson} and multiply \eqref{eq:heat} with a weight function $w$ and integrate over the whole domain:
\begin{equation}
\label{eq:heat_weak1}
\int_{\Omega} w \grad \cdot (\kappa \grad T) d\Omega +
\int_{\Omega} w \left( P_J-C_v \dt{T} \right) d\Omega = 0 \period
\end{equation}

The properties of differentials allow expressing \eqref{eq:heat_weak1} as
\begin{equation}
\label{eq:heat_weak2}
\begin{split}
\int_{\Omega} \grad \cdot ( w\kappa \grad T) d\Omega -
\int_{\Omega} \grad w \cdot \kappa \grad T d\Omega \\+
\int_{\Omega} w \left( P_J-C_v \dt{T} \right) d\Omega = 0 \period
\end{split}
\end{equation}

By means of Green’s theorem, we remove the double differentiation of $T$ from the first term of \eqref{eq:heat_weak2},
\begin{equation}
\label{eq:heat_weak3}
\begin{split}
\oint_{\Gamma} w \kappa \grad T \cdot \vec{n} d\Gamma -
\int_{\Omega} \grad w \cdot \kappa \grad T d\Omega \\+
\int_{\Omega} w \left( P_J-C_v \dt{T} \right) d\Omega = 0 \period
\end{split}
\end{equation}

After plugging BCs \eqref{eq:heat_bc} into the previous formula and rearranging terms, we obtain the weak form of eq. \eqref{eq:heat}
\begin{equation}
\label{eq:heat_weak}
\begin{split}
\int_{\Omega_2} wC_v \dt{T} d\Omega +
\int_{\Omega_2} \kappa \grad w \cdot \grad T d\Omega \\=
\int_{\Omega_2} wP_J d\Omega + \int_{\Gamma_3} wP_N d\Gamma \period
\end{split}
\end{equation}

To express the weak form as a matrix equation, we follow the procedure from appendix \ref{sec:weak_poisson} and equalize the weight function with shape functions and expand the temperature as a linear combination of them.
In that way we separate spatial and temporal parts in $T$, and eq. \eqref{eq:heat_weak} becomes
\begin{equation}
\label{eq:heat_matlong}
\begin{split}
\int_{\Omega_2} C_v N_i \sum_{j=1}^{n} \left(N_j \dt{T_j} \right) d\Omega +
\int_{\Omega_2} \kappa \sum_{j=1}^{n} (\grad N_i \cdot \grad N_j T_j) d\Omega \\=
\int_{\Omega_2} N_iP_J d\Omega + \int_{\Gamma_3} N_iP_N d\Gamma \comma
\end{split}    
\end{equation}
which looks in matrix form as
\begin{equation}
\label{eq:heat_mat}
\vecb{C} \cdot \dt{\vecb{T}} + \vecb{K} \cdot \vecb{T} = \vecb{f} \comma
\end{equation}
where
\begin{equation*}
\begin{aligned}
C_{ij} &= \int_{\Omega_2} C_vN_iN_j d\Omega \comma\\
K_{ij} &= \int_{\Omega_2} \kappa \grad N_i \cdot \grad N_j d\Omega \comma \\
f_i &= \int_{\Omega_2} N_i P_J d\Omega + \int_{\Gamma_3} N_i P_N d\Gamma \period
\end{aligned}
\end{equation*}

To solve semidiscrete eq. \eqref{eq:heat_mat}, the time also must be discretized.
It can be done by first approximating
\begin{equation}
\label{eq:heat_timediscr}
\dt{\vecb{T}^n} \approx
\frac{\vecb{T}^{n+1} - \vecb{T}^n}{\Delta t}
\end{equation}
and then introducing parameter $\Theta \in [0,1]$, so that
\begin{equation}
\label{eq:heat_theta1}
\vecb{T}^{n+\Theta} =
\Theta \vecb{T}^{n+1} + (1-\Theta)\vecb{T}^n \period
\end{equation}
After combining \eqref{eq:heat_timediscr} and \eqref{eq:heat_theta1} together, we get a $\Theta$-scheme for solving eq. \eqref{eq:heat_mat}:
\begin{equation}
\label{eq:heat_theta2}
\begin{split}
\left( \frac{\vecb{C}^n}{\Delta t} + \Theta \vecb{K}^{n+1} \right)
\vecb{T}^{n+1} =\\ 
\left[ \frac{\vecb{C}^n}{\Delta t} - (1-\Theta)\vecb{K}^n \right] 
\vecb{T}^n +
\Theta \vecb{f}^{n+1} + (1-\Theta)\vecb{f}^n \period
\end{split}
\end{equation}

To handle the nonlinearity, we assume in this formula, that thermal and electric conductivities are weakly temperature dependent, i.e., $\kappa(t_{n+1}) \approx \kappa(t_n)$ and $\sigma(t_{n+1}) \approx \sigma(t_n)$.
In the case of $\Theta=0$, $\Theta=0.5$, and $\Theta = 1$, we get the explicit Euler, Crank-Nicolson and implicit Euler scheme, respectively.
In our simulations we use the implicit Euler scheme while solving eq. \eqref{eq:heat}.
This choice helps to quickly diffuse high-frequency noise in the temperature distribution that is introduced while mapping temperatures during the mesh rebuild.
Although the implicit Euler scheme shows a lower order of convergence than the Crank-Nicolson method, i.e., smaller $\Delta t$ needs to be used \cite{lewis_fundamentals_2004}, in our simulations the small time step is determined by MD, and therefore this shortcoming can be safely ignored in favor of more preferential diffusive properties. 

\vspace{10pt}
\section{Distributing injected superparticles}
\label{sec:rnd_injection}
\begin{figure}[h!]
    \includegraphics[scale=0.7]{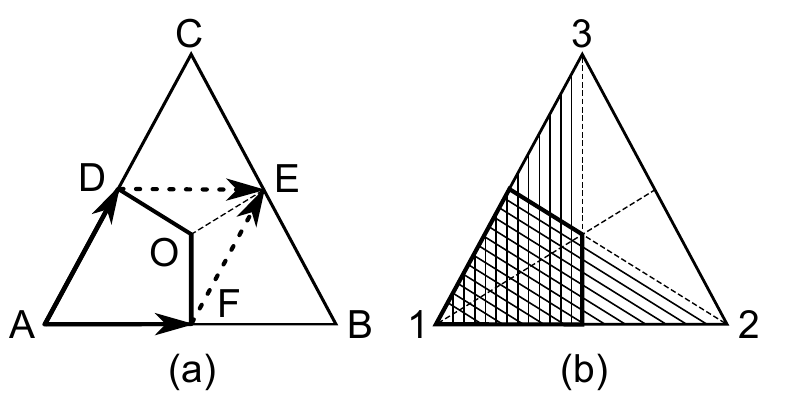}
    \caption{Generation of a superparticle into a quadrangle, which in turn is located inside a triangle.}
    \label{fig:locate_rnd_point}
\end{figure}

The injected superparticles must be distributed randomly and uniformly on the quadrangle.
This can be done in a straightforward way, by noting that the mesh quadrangles are built by connecting the centroids of the mesh edges and the centroids of triangles (see fig. \ref{fig:locate_rnd_point}a).
The coordinates of random point inside a parallelogram $ADEF$ are given by
\begin{equation}
    \label{eq:rnd_point}
    \lvec{r} = \lvec{A} + R_1\lvec{AD} + R_2\lvec{AF} \comma
\end{equation}
where $R_1, R_2 \in [0,1]$ are uniform random numbers.
The point $\vec{r}$, however, might not lie inside the quadrangle $ADOF$, as its area is smaller than the area of the parallelogram, $S_{ADOF} = \frac{1}{3} S_{ABC} < S_{ADEF} = \frac{1}{2} S_{ABC}$.
Therefore its location in the quadrangle must be checked, and if $\vec{r}$ turns to be out of $ADEF$, a new point should be generated and the check procedure repeated.

The location of the point can be determined with barycentric coordinates \cite{vince_barycentric_2014}.
Denoting $\lambda_1$, $\lambda_2$ and $\lambda_3$ as the barycentric coordinates of point $\vec{r}$ with respect to the first, second and third node of a triangle (fig. \ref{fig:locate_rnd_point}b), $\vec{r}$ is located in the left region of the triangle (vertical hatching) if $\lambda_1 \geq \lambda_2$ and in the bottom diagonal region (diagonal hatching), if $\lambda_1 \geq \lambda_3$.

\vspace{10pt}
\section{Collisions between superparticles of the same type}
\label{sec:collisions}

The Monte Carlo binary collision model can be described in the following steps:
\begin{enumerate}
    \item Divide SPs into groups in such a way that all particles within the group are located in the same cell.
    \item Pair the SPs in the group in a random way.
    \item Collide each pair elastically by assuming a normal distribution of the scattering angles.
\end{enumerate}

The pairing of the SPs is implemented here as described in \cite{takizuka_binary_1977}.
To collide elastically SP pairs, we use the statistical model as follows. 
By assuming energy and momentum conservation, the velocities of paired particles of identical mass $m$ and charge $q$ before and after the collision, $\vec{u}_{1,2}$ and $\vec{v}_{1,2}$, are written as
\begin{equation}
\label{eq:vel12}
\vec{v}_{1,2} = \vec{u}_{1,2} \pm 0.5\Delta \vec{v} \period
\end{equation}

The velocity change $\Delta \vec{v}$ is calculated from the scattering of the relative velocity,
\begin{equation}
\label{eq:scat_vel}
\Delta \vec{v} = \vecb{M}(\theta, \phi) \cdot \left( \vec{u}_2 - \vec{u}_1 \right) \comma
\end{equation}
where the rotation matrix $\vecb{M}$ is expressed as
\begin{equation}
\label{eq:rot_mat}
\vecb{M}(\theta, \phi) =
\begin{pmatrix*}[l]
\cos\theta-1 &a &b_x\\ 
-a &\cos\theta-1 &b_y\\
-b_x &-b_y &\cos\theta-1
\end{pmatrix*} \comma
\end{equation}
where $ a = \sin\theta \sin\phi \frac{v}{v_\perp}$, $b_{x,y} = \sin\theta \cos\phi \frac{v_{x,y}}{v_\perp}$ and $v_\perp = \sqrt{v_x^2 + v_y^2}$.
The Monte Carlo technique comes into play while calculating the azimuth and scattering angle $\phi$~and~$\theta$.
To calculate the first, we generate a uniform random number in the range of $[0, 2\pi)$.
Scattering angles~$\theta$, however, need to be distributed normally and their contribution to matrix \eqref{eq:rot_mat} are found from
\begin{equation}
\label{eq:sin_theta}
\begin{aligned}
\sin\theta &= \frac{2\delta}{1+\delta^2} \comma\\
1-\cos\theta &= \frac{2\delta^2}{1+\delta^2} \comma
\end{aligned}    
\end{equation}
where $\delta \equiv \tan\theta$ is a random number generated according to the Gauss distribution.
The mean of that distribution $\langle \delta \rangle = 0$ and the variance for a system of SPs with identical charge and mass
\begin{equation}
\label{eq:variance}
\langle \delta^2 \rangle = \frac{q^4 n \Lambda \Delta t}{2 \pi \epsilon_0^2 m^2 v^3 V_{cell}} \comma
\end{equation}
where $n$ is the number of particles in the cell, $V_{cell}$ the cell volume, $\epsilon_0$ the vacuum permittivity and $\Lambda=13.0$ the Landau logarithm \cite{krall_principles_1973}.
In the case that there is odd number of SPs in the cell, the first three pairs in that cell should be collided with a variance of $\langle \delta^2 \rangle^{'} = 0.5\langle \delta^2 \rangle$ \cite{takizuka_binary_1977}.

\vfill
\pagebreak
\bibliography{bib/zotero,bib/abbreviated}

\begin{thebibliography}{57}%
\makeatletter
\providecommand \@ifxundefined [1]{%
 \@ifx{#1\undefined}
}%
\providecommand \@ifnum [1]{%
 \ifnum #1\expandafter \@firstoftwo
 \else \expandafter \@secondoftwo
 \fi
}%
\providecommand \@ifx [1]{%
 \ifx #1\expandafter \@firstoftwo
 \else \expandafter \@secondoftwo
 \fi
}%
\providecommand \natexlab [1]{#1}%
\providecommand \enquote  [1]{``#1''}%
\providecommand \bibnamefont  [1]{#1}%
\providecommand \bibfnamefont [1]{#1}%
\providecommand \citenamefont [1]{#1}%
\providecommand \href@noop [0]{\@secondoftwo}%
\providecommand \href [0]{\begingroup \@sanitize@url \@href}%
\providecommand \@href[1]{\@@startlink{#1}\@@href}%
\providecommand \@@href[1]{\endgroup#1\@@endlink}%
\providecommand \@sanitize@url [0]{\catcode `\\12\catcode `\$12\catcode
  `\&12\catcode `\#12\catcode `\^12\catcode `\_12\catcode `\%12\relax}%
\providecommand \@@startlink[1]{}%
\providecommand \@@endlink[0]{}%
\providecommand \url  [0]{\begingroup\@sanitize@url \@url }%
\providecommand \@url [1]{\endgroup\@href {#1}{\urlprefix }}%
\providecommand \urlprefix  [0]{URL }%
\providecommand \Eprint [0]{\href }%
\providecommand \doibase [0]{http://dx.doi.org/}%
\providecommand \selectlanguage [0]{\@gobble}%
\providecommand \bibinfo  [0]{\@secondoftwo}%
\providecommand \bibfield  [0]{\@secondoftwo}%
\providecommand \translation [1]{[#1]}%
\providecommand \BibitemOpen [0]{}%
\providecommand \bibitemStop [0]{}%
\providecommand \bibitemNoStop [0]{.\EOS\space}%
\providecommand \EOS [0]{\spacefactor3000\relax}%
\providecommand \BibitemShut  [1]{\csname bibitem#1\endcsname}%
\let\auto@bib@innerbib\@empty
%</preamble>
\bibitem [{\citenamefont {Slade}(2007)}]{slade2007}%
  \BibitemOpen
  \bibfield  {author} {\bibinfo {author} {\bibfnamefont {P.}~\bibnamefont
  {Slade}},\ }\href {https://books.google.fi/books?id=uJT4mENgMbQC} {\emph
  {\bibinfo {title} {The Vacuum Interrupter: Theory, Design, and
  Application}}}\ (\bibinfo  {publisher} {CRC Press},\ \bibinfo {year}
  {2007})\BibitemShut {NoStop}%
\bibitem [{\citenamefont {KNAW}(2019)}]{knaw_smart*light:_2019}%
  \BibitemOpen
  \bibfield  {author} {\bibinfo {author} {\bibnamefont {KNAW}},\ }\href
  {https://www.knaw.nl/shared/resources/adviezen/EervolKNAWAgendaSmartLight.pdf}
  {\enquote {\bibinfo {title} {Smart*{Light}: a {Dutch} table-top synchrotron
  light source},}\ } (\bibinfo {year} {2019})\BibitemShut {NoStop}%
\bibitem [{\citenamefont {McCracken}(1980)}]{McCracken1980}%
  \BibitemOpen
  \bibfield  {author} {\bibinfo {author} {\bibfnamefont {G.}~\bibnamefont
  {McCracken}},\ }\href {\doibase 10.1016/0022-3115(80)90299-8} {\bibfield
  {journal} {\bibinfo  {journal} {J. Nucl. Mater.}\ }\textbf {\bibinfo {volume}
  {93}},\ \bibinfo {pages} {3 } (\bibinfo {year} {1980})}\BibitemShut {NoStop}%
\bibitem [{\citenamefont {{CERN}}(2018)}]{cern_compact_2018}%
  \BibitemOpen
  \bibfield  {author} {\bibinfo {author} {\bibnamefont {{CERN}}},\ }\href
  {\doibase 10.23731/cyrm-2018-002} {\bibfield  {journal} {\bibinfo  {journal}
  {CERN Yellow Reports}\ } (\bibinfo {year} {2018}),\
  10.23731/cyrm-2018-002}\BibitemShut {NoStop}%
\bibitem [{\citenamefont {Dyke}\ and\ \citenamefont
  {Trolan}(1953)}]{Dyke1953Arc}%
  \BibitemOpen
  \bibfield  {author} {\bibinfo {author} {\bibfnamefont {W.}~\bibnamefont
  {Dyke}}\ and\ \bibinfo {author} {\bibfnamefont {J.}~\bibnamefont {Trolan}},\
  }\href {\doibase 10.1103/PhysRev.89.799} {\bibfield  {journal} {\bibinfo
  {journal} {Phys. Rev.}\ }\textbf {\bibinfo {volume} {89}},\ \bibinfo {pages}
  {799} (\bibinfo {year} {1953})}\BibitemShut {NoStop}%
\bibitem [{\citenamefont {Dyke}\ \emph {et~al.}(1953)\citenamefont {Dyke},
  \citenamefont {Trolan}, \citenamefont {Martin},\ and\ \citenamefont
  {Barbour}}]{Dyke1953I}%
  \BibitemOpen
  \bibfield  {author} {\bibinfo {author} {\bibfnamefont {W.}~\bibnamefont
  {Dyke}}, \bibinfo {author} {\bibfnamefont {J.}~\bibnamefont {Trolan}},
  \bibinfo {author} {\bibfnamefont {E.}~\bibnamefont {Martin}}, \ and\ \bibinfo
  {author} {\bibfnamefont {J.}~\bibnamefont {Barbour}},\ }\href {\doibase
  10.1103/PhysRev.91.1043} {\bibfield  {journal} {\bibinfo  {journal} {Phys.
  Rev.}\ }\textbf {\bibinfo {volume} {91}},\ \bibinfo {pages} {1043} (\bibinfo
  {year} {1953})}\BibitemShut {NoStop}%
\bibitem [{\citenamefont {Hartmann}\ and\ \citenamefont
  {Gundersen}(1988)}]{Anders_PRL88}%
  \BibitemOpen
  \bibfield  {author} {\bibinfo {author} {\bibfnamefont {W.}~\bibnamefont
  {Hartmann}}\ and\ \bibinfo {author} {\bibfnamefont {M.~A.}\ \bibnamefont
  {Gundersen}},\ }\href {\doibase 10.1103/PhysRevLett.60.2371} {\bibfield
  {journal} {\bibinfo  {journal} {Phys. Rev. Lett.}\ }\textbf {\bibinfo
  {volume} {60}},\ \bibinfo {pages} {2371} (\bibinfo {year}
  {1988})}\BibitemShut {NoStop}%
\bibitem [{\citenamefont {Anders}(2008)}]{Anders}%
  \BibitemOpen
  \bibfield  {author} {\bibinfo {author} {\bibfnamefont {A.}~\bibnamefont
  {Anders}},\ }\href {\doibase 10.1007/978-0-387-79108-1} {\emph {\bibinfo
  {title} {Cathodic Arcs}}},\ \bibinfo {edition} {1st}\ ed.,\ Vol.~\bibinfo
  {volume} {3}\ (\bibinfo  {publisher} {Springer-Verlag New York},\ \bibinfo
  {year} {2008})\BibitemShut {NoStop}%
\bibitem [{\citenamefont {Mesyats}(1993)}]{mesyats1993ectons}%
  \BibitemOpen
  \bibfield  {author} {\bibinfo {author} {\bibfnamefont {G.}~\bibnamefont
  {Mesyats}},\ }\href@noop {} {\bibfield  {journal} {\bibinfo  {journal} {JETP
  Lett.}\ }\textbf {\bibinfo {volume} {57}},\ \bibinfo {pages} {95} (\bibinfo
  {year} {1993})}\BibitemShut {NoStop}%
\bibitem [{\citenamefont {Mesyats}(2005)}]{Mesyats2005}%
  \BibitemOpen
  \bibfield  {author} {\bibinfo {author} {\bibfnamefont {G.~A.}\ \bibnamefont
  {Mesyats}},\ }\href {http://stacks.iop.org/0741-3335/47/i=5A/a=010}
  {\bibfield  {journal} {\bibinfo  {journal} {Plasma Phys. Controlled Fusion}\
  }\textbf {\bibinfo {volume} {47}},\ \bibinfo {pages} {A109} (\bibinfo {year}
  {2005})}\BibitemShut {NoStop}%
\bibitem [{\citenamefont {Timko}\ \emph {et~al.}(2015)\citenamefont {Timko},
  \citenamefont {Ness~Sjobak}, \citenamefont {Mether}, \citenamefont
  {Calatroni}, \citenamefont {Djurabekova}, \citenamefont {Matyash},
  \citenamefont {Nordlund}, \citenamefont {Schneider},\ and\ \citenamefont
  {Wuensch}}]{timko_field_2015}%
  \BibitemOpen
  \bibfield  {author} {\bibinfo {author} {\bibfnamefont {H.}~\bibnamefont
  {Timko}}, \bibinfo {author} {\bibfnamefont {K.}~\bibnamefont {Ness~Sjobak}},
  \bibinfo {author} {\bibfnamefont {L.}~\bibnamefont {Mether}}, \bibinfo
  {author} {\bibfnamefont {S.}~\bibnamefont {Calatroni}}, \bibinfo {author}
  {\bibfnamefont {F.}~\bibnamefont {Djurabekova}}, \bibinfo {author}
  {\bibfnamefont {K.}~\bibnamefont {Matyash}}, \bibinfo {author} {\bibfnamefont
  {K.}~\bibnamefont {Nordlund}}, \bibinfo {author} {\bibfnamefont
  {R.}~\bibnamefont {Schneider}}, \ and\ \bibinfo {author} {\bibfnamefont
  {W.}~\bibnamefont {Wuensch}},\ }\href {\doibase 10.1002/ctpp.201400069}
  {\bibfield  {journal} {\bibinfo  {journal} {Contributions to Plasma Physics}\
  }\textbf {\bibinfo {volume} {55}},\ \bibinfo {pages} {299} (\bibinfo {year}
  {2015})}\BibitemShut {NoStop}%
\bibitem [{\citenamefont {Kyritsakis}\ \emph {et~al.}(2018)\citenamefont
  {Kyritsakis}, \citenamefont {Veske}, \citenamefont {Eimre}, \citenamefont
  {Zadin},\ and\ \citenamefont {Djurabekova}}]{kyritsakis_thermal_2018}%
  \BibitemOpen
  \bibfield  {author} {\bibinfo {author} {\bibfnamefont {A.}~\bibnamefont
  {Kyritsakis}}, \bibinfo {author} {\bibfnamefont {M.}~\bibnamefont {Veske}},
  \bibinfo {author} {\bibfnamefont {K.}~\bibnamefont {Eimre}}, \bibinfo
  {author} {\bibfnamefont {V.}~\bibnamefont {Zadin}}, \ and\ \bibinfo {author}
  {\bibfnamefont {F.}~\bibnamefont {Djurabekova}},\ }\href {\doibase
  10.1088/1361-6463/aac03b} {\bibfield  {journal} {\bibinfo  {journal} {Journal
  of Physics D: Applied Physics}\ }\textbf {\bibinfo {volume} {51}},\ \bibinfo
  {pages} {225203} (\bibinfo {year} {2018})}\BibitemShut {NoStop}%
\bibitem [{\citenamefont {Forbes}(2008)}]{forbes_exact_2008}%
  \BibitemOpen
  \bibfield  {author} {\bibinfo {author} {\bibfnamefont {R.~G.}\ \bibnamefont
  {Forbes}},\ }\href {\doibase 10.1063/1.2996005} {\bibfield  {journal}
  {\bibinfo  {journal} {Journal of Applied Physics}\ }\textbf {\bibinfo
  {volume} {104}},\ \bibinfo {pages} {084303} (\bibinfo {year}
  {2008})}\BibitemShut {NoStop}%
\bibitem [{\citenamefont {Uimanov}(2011)}]{Uimanov2011}%
  \BibitemOpen
  \bibfield  {author} {\bibinfo {author} {\bibfnamefont {I.~V.}\ \bibnamefont
  {Uimanov}},\ }\href {\doibase 10.1109/TDEI.2011.5931082} {\bibfield
  {journal} {\bibinfo  {journal} {IEEE Transactions on Dielectrics and
  Electrical Insulation}\ }\textbf {\bibinfo {volume} {18}},\ \bibinfo {pages}
  {924} (\bibinfo {year} {2011})}\BibitemShut {NoStop}%
\bibitem [{\citenamefont {Torfason}\ \emph {et~al.}(2015)\citenamefont
  {Torfason}, \citenamefont {Valfells},\ and\ \citenamefont
  {Manolescu}}]{torfason_molecular_2015}%
  \BibitemOpen
  \bibfield  {author} {\bibinfo {author} {\bibfnamefont {K.}~\bibnamefont
  {Torfason}}, \bibinfo {author} {\bibfnamefont {A.}~\bibnamefont {Valfells}},
  \ and\ \bibinfo {author} {\bibfnamefont {A.}~\bibnamefont {Manolescu}},\
  }\href {\doibase 10.1063/1.4914855} {\bibfield  {journal} {\bibinfo
  {journal} {Physics of Plasmas}\ }\textbf {\bibinfo {volume} {22}},\ \bibinfo
  {pages} {033109} (\bibinfo {year} {2015})}\BibitemShut {NoStop}%
\bibitem [{\citenamefont {Torfason}\ \emph {et~al.}(2016)\citenamefont
  {Torfason}, \citenamefont {Valfells},\ and\ \citenamefont
  {Manolescu}}]{torfason_molecular_2016}%
  \BibitemOpen
  \bibfield  {author} {\bibinfo {author} {\bibfnamefont {K.}~\bibnamefont
  {Torfason}}, \bibinfo {author} {\bibfnamefont {A.}~\bibnamefont {Valfells}},
  \ and\ \bibinfo {author} {\bibfnamefont {A.}~\bibnamefont {Manolescu}},\
  }\href {\doibase 10.1063/1.4972821} {\bibfield  {journal} {\bibinfo
  {journal} {Physics of Plasmas}\ }\textbf {\bibinfo {volume} {23}},\ \bibinfo
  {pages} {123119} (\bibinfo {year} {2016})}\BibitemShut {NoStop}%
\bibitem [{\citenamefont {Tskhakaya}\ \emph {et~al.}(2007)\citenamefont
  {Tskhakaya}, \citenamefont {Matyash}, \citenamefont {Schneider},\ and\
  \citenamefont {Taccogna}}]{tskhakaya_particle--cell_2007}%
  \BibitemOpen
  \bibfield  {author} {\bibinfo {author} {\bibfnamefont {D.}~\bibnamefont
  {Tskhakaya}}, \bibinfo {author} {\bibfnamefont {K.}~\bibnamefont {Matyash}},
  \bibinfo {author} {\bibfnamefont {R.}~\bibnamefont {Schneider}}, \ and\
  \bibinfo {author} {\bibfnamefont {F.}~\bibnamefont {Taccogna}},\ }\href
  {\doibase 10.1002/ctpp.200710072} {\bibfield  {journal} {\bibinfo  {journal}
  {Contributions to Plasma Physics}\ }\textbf {\bibinfo {volume} {47}},\
  \bibinfo {pages} {563} (\bibinfo {year} {2007})}\BibitemShut {NoStop}%
\bibitem [{\citenamefont {Timko}(2011)}]{timko_modelling_2011}%
  \BibitemOpen
  \bibfield  {author} {\bibinfo {author} {\bibfnamefont {H.}~\bibnamefont
  {Timko}},\ }\emph {\bibinfo {title} {Modelling vacuum arcs : from plasma
  initiation to surface interactions}},\ \href
  {https://helda.helsinki.fi/handle/10138/28262} {Ph.D. thesis},\ \bibinfo
  {school} {University of Helsinki}, \bibinfo {address} {Helsinki} (\bibinfo
  {year} {2011})\BibitemShut {NoStop}%
\bibitem [{\citenamefont {Veske}\ \emph {et~al.}(2018)\citenamefont {Veske},
  \citenamefont {Kyritsakis}, \citenamefont {Eimre}, \citenamefont {Zadin},
  \citenamefont {Aabloo},\ and\ \citenamefont
  {Djurabekova}}]{veske_dynamic_2018}%
  \BibitemOpen
  \bibfield  {author} {\bibinfo {author} {\bibfnamefont {M.}~\bibnamefont
  {Veske}}, \bibinfo {author} {\bibfnamefont {A.}~\bibnamefont {Kyritsakis}},
  \bibinfo {author} {\bibfnamefont {K.}~\bibnamefont {Eimre}}, \bibinfo
  {author} {\bibfnamefont {V.}~\bibnamefont {Zadin}}, \bibinfo {author}
  {\bibfnamefont {A.}~\bibnamefont {Aabloo}}, \ and\ \bibinfo {author}
  {\bibfnamefont {F.}~\bibnamefont {Djurabekova}},\ }\href {\doibase
  10.1016/j.jcp.2018.04.031} {\bibfield  {journal} {\bibinfo  {journal}
  {Journal of Computational Physics}\ }\textbf {\bibinfo {volume} {367}},\
  \bibinfo {pages} {279} (\bibinfo {year} {2018})}\BibitemShut {NoStop}%
\bibitem [{\citenamefont {Birdsall}\ and\ \citenamefont
  {Langdon}(1985)}]{birdsall_plasma_1985}%
  \BibitemOpen
  \bibfield  {author} {\bibinfo {author} {\bibfnamefont {C.~K.}\ \bibnamefont
  {Birdsall}}\ and\ \bibinfo {author} {\bibfnamefont {A.~B.}\ \bibnamefont
  {Langdon}},\ }\href@noop {} {\emph {\bibinfo {title} {Plasma physics via
  computer simulation}}}\ (\bibinfo  {publisher} {McGraw-Hill},\ \bibinfo
  {address} {New York},\ \bibinfo {year} {1985})\BibitemShut {NoStop}%
\bibitem [{\citenamefont {Buneman}(1959)}]{buneman_dissipation_1959}%
  \BibitemOpen
  \bibfield  {author} {\bibinfo {author} {\bibfnamefont {O.}~\bibnamefont
  {Buneman}},\ }\href {\doibase 10.1103/PhysRev.115.503} {\bibfield  {journal}
  {\bibinfo  {journal} {Physical Review}\ }\textbf {\bibinfo {volume} {115}},\
  \bibinfo {pages} {503} (\bibinfo {year} {1959})}\BibitemShut {NoStop}%
\bibitem [{\citenamefont {Dawson}(1962)}]{dawson_one-dimensional_1962}%
  \BibitemOpen
  \bibfield  {author} {\bibinfo {author} {\bibfnamefont {J.}~\bibnamefont
  {Dawson}},\ }\href {\doibase 10.1063/1.1706638} {\bibfield  {journal}
  {\bibinfo  {journal} {Physics of Fluids}\ }\textbf {\bibinfo {volume} {5}},\
  \bibinfo {pages} {445} (\bibinfo {year} {1962})}\BibitemShut {NoStop}%
\bibitem [{\citenamefont {Brooks}\ and\ \citenamefont
  {Naujoks}(2000)}]{brooks_sheath_2000}%
  \BibitemOpen
  \bibfield  {author} {\bibinfo {author} {\bibfnamefont {J.~N.}\ \bibnamefont
  {Brooks}}\ and\ \bibinfo {author} {\bibfnamefont {D.}~\bibnamefont
  {Naujoks}},\ }\href {\doibase 10.1063/1.874097} {\bibfield  {journal}
  {\bibinfo  {journal} {Physics of Plasmas}\ }\textbf {\bibinfo {volume} {7}},\
  \bibinfo {pages} {2565} (\bibinfo {year} {2000})}\BibitemShut {NoStop}%
\bibitem [{\citenamefont {Taccogna}\ \emph
  {et~al.}(2004{\natexlab{a}})\citenamefont {Taccogna}, \citenamefont {Longo},\
  and\ \citenamefont {Capitelli}}]{taccogna_effects_2004}%
  \BibitemOpen
  \bibfield  {author} {\bibinfo {author} {\bibfnamefont {F.}~\bibnamefont
  {Taccogna}}, \bibinfo {author} {\bibfnamefont {S.}~\bibnamefont {Longo}}, \
  and\ \bibinfo {author} {\bibfnamefont {M.}~\bibnamefont {Capitelli}},\ }\href
  {\doibase 10.1016/j.vacuum.2003.12.039} {\bibfield  {journal} {\bibinfo
  {journal} {Vacuum}\ }\textbf {\bibinfo {volume} {73}},\ \bibinfo {pages} {89}
  (\bibinfo {year} {2004}{\natexlab{a}})}\BibitemShut {NoStop}%
\bibitem [{\citenamefont {Taccogna}\ \emph {et~al.}(2007)\citenamefont
  {Taccogna}, \citenamefont {Schneider}, \citenamefont {Matyash}, \citenamefont
  {Longo}, \citenamefont {Capitelli},\ and\ \citenamefont
  {Tskhakaya}}]{taccogna_negative_2007}%
  \BibitemOpen
  \bibfield  {author} {\bibinfo {author} {\bibfnamefont {F.}~\bibnamefont
  {Taccogna}}, \bibinfo {author} {\bibfnamefont {R.}~\bibnamefont {Schneider}},
  \bibinfo {author} {\bibfnamefont {K.}~\bibnamefont {Matyash}}, \bibinfo
  {author} {\bibfnamefont {S.}~\bibnamefont {Longo}}, \bibinfo {author}
  {\bibfnamefont {M.}~\bibnamefont {Capitelli}}, \ and\ \bibinfo {author}
  {\bibfnamefont {D.}~\bibnamefont {Tskhakaya}},\ }\href {\doibase
  10.1016/j.jnucmat.2007.01.039} {\bibfield  {journal} {\bibinfo  {journal}
  {Journal of Nuclear Materials}\ }\textbf {\bibinfo {volume} {363-365}},\
  \bibinfo {pages} {437} (\bibinfo {year} {2007})}\BibitemShut {NoStop}%
\bibitem [{\citenamefont {Taccogna}\ \emph
  {et~al.}(2004{\natexlab{b}})\citenamefont {Taccogna}, \citenamefont {Longo},\
  and\ \citenamefont {Capitelli}}]{taccogna_plasma-surface_2004}%
  \BibitemOpen
  \bibfield  {author} {\bibinfo {author} {\bibfnamefont {F.}~\bibnamefont
  {Taccogna}}, \bibinfo {author} {\bibfnamefont {S.}~\bibnamefont {Longo}}, \
  and\ \bibinfo {author} {\bibfnamefont {M.}~\bibnamefont {Capitelli}},\ }\href
  {\doibase 10.1063/1.1647567} {\bibfield  {journal} {\bibinfo  {journal}
  {Physics of Plasmas}\ }\textbf {\bibinfo {volume} {11}},\ \bibinfo {pages}
  {1220} (\bibinfo {year} {2004}{\natexlab{b}})}\BibitemShut {NoStop}%
\bibitem [{\citenamefont {Taccogna}\ \emph {et~al.}(2005)\citenamefont
  {Taccogna}, \citenamefont {Longo},\ and\ \citenamefont
  {Capitelli}}]{taccogna_plasma_2005}%
  \BibitemOpen
  \bibfield  {author} {\bibinfo {author} {\bibfnamefont {F.}~\bibnamefont
  {Taccogna}}, \bibinfo {author} {\bibfnamefont {S.}~\bibnamefont {Longo}}, \
  and\ \bibinfo {author} {\bibfnamefont {M.}~\bibnamefont {Capitelli}},\ }\href
  {\doibase 10.1063/1.2015257} {\bibfield  {journal} {\bibinfo  {journal}
  {Physics of Plasmas}\ }\textbf {\bibinfo {volume} {12}},\ \bibinfo {pages}
  {093506} (\bibinfo {year} {2005})}\BibitemShut {NoStop}%
\bibitem [{\citenamefont {Kyritsakis}\ and\ \citenamefont
  {Djurabekova}(2017)}]{kyritsakis_general_2017}%
  \BibitemOpen
  \bibfield  {author} {\bibinfo {author} {\bibfnamefont {A.}~\bibnamefont
  {Kyritsakis}}\ and\ \bibinfo {author} {\bibfnamefont {F.}~\bibnamefont
  {Djurabekova}},\ }\href {\doibase 10.1016/j.commatsci.2016.11.010} {\bibfield
   {journal} {\bibinfo  {journal} {Computational Materials Science}\ }\textbf
  {\bibinfo {volume} {128}},\ \bibinfo {pages} {15} (\bibinfo {year}
  {2017})}\BibitemShut {NoStop}%
\bibitem [{\citenamefont {Takizuka}\ and\ \citenamefont
  {Abe}(1977)}]{takizuka_binary_1977}%
  \BibitemOpen
  \bibfield  {author} {\bibinfo {author} {\bibfnamefont {T.}~\bibnamefont
  {Takizuka}}\ and\ \bibinfo {author} {\bibfnamefont {H.}~\bibnamefont {Abe}},\
  }\href {\doibase 10.1016/0021-9991(77)90099-7} {\bibfield  {journal}
  {\bibinfo  {journal} {Journal of Computational Physics}\ }\textbf {\bibinfo
  {volume} {25}},\ \bibinfo {pages} {205} (\bibinfo {year} {1977})}\BibitemShut
  {NoStop}%
\bibitem [{\citenamefont {Landau}\ and\ \citenamefont
  {Lifshitz}(1984)}]{landau_electrodynamics_1984}%
  \BibitemOpen
  \bibfield  {author} {\bibinfo {author} {\bibfnamefont {L.~D.}\ \bibnamefont
  {Landau}}\ and\ \bibinfo {author} {\bibfnamefont {E.~M.}\ \bibnamefont
  {Lifshitz}},\ }\href@noop {} {\emph {\bibinfo {title} {Electrodynamics of
  {Continuous} {Media}}}},\ Vol.~\bibinfo {volume} {8}\ (\bibinfo  {publisher}
  {Pergamon},\ \bibinfo {year} {1984})\BibitemShut {NoStop}%
\bibitem [{\citenamefont {Djurabekova}\ \emph {et~al.}(2011)\citenamefont
  {Djurabekova}, \citenamefont {Parviainen}, \citenamefont {Pohjonen},\ and\
  \citenamefont {Nordlund}}]{djurabekova_atomistic_2011}%
  \BibitemOpen
  \bibfield  {author} {\bibinfo {author} {\bibfnamefont {F.}~\bibnamefont
  {Djurabekova}}, \bibinfo {author} {\bibfnamefont {S.}~\bibnamefont
  {Parviainen}}, \bibinfo {author} {\bibfnamefont {A.}~\bibnamefont
  {Pohjonen}}, \ and\ \bibinfo {author} {\bibfnamefont {K.}~\bibnamefont
  {Nordlund}},\ }\href {\doibase 10.1103/PhysRevE.83.026704} {\bibfield
  {journal} {\bibinfo  {journal} {Physical Review E}\ }\textbf {\bibinfo
  {volume} {83}},\ \bibinfo {pages} {026704} (\bibinfo {year}
  {2011})}\BibitemShut {NoStop}%
\bibitem [{\citenamefont {Si}(2015)}]{si_tetgen_2015}%
  \BibitemOpen
  \bibfield  {author} {\bibinfo {author} {\bibfnamefont {H.}~\bibnamefont
  {Si}},\ }\href {\doibase 10.1145/2629697} {\bibfield  {journal} {\bibinfo
  {journal} {ACM Transactions on Mathematical Software}\ }\textbf {\bibinfo
  {volume} {41}},\ \bibinfo {pages} {36} (\bibinfo {year} {2015})}\BibitemShut
  {NoStop}%
\bibitem [{\citenamefont {Parviainen}\ \emph {et~al.}(2011)\citenamefont
  {Parviainen}, \citenamefont {Djurabekova}, \citenamefont {Timko},\ and\
  \citenamefont {Nordlund}}]{parviainen_electronic_2011}%
  \BibitemOpen
  \bibfield  {author} {\bibinfo {author} {\bibfnamefont {S.}~\bibnamefont
  {Parviainen}}, \bibinfo {author} {\bibfnamefont {F.}~\bibnamefont
  {Djurabekova}}, \bibinfo {author} {\bibfnamefont {H.}~\bibnamefont {Timko}},
  \ and\ \bibinfo {author} {\bibfnamefont {K.}~\bibnamefont {Nordlund}},\
  }\href {\doibase 10.1016/j.commatsci.2011.02.010} {\bibfield  {journal}
  {\bibinfo  {journal} {Computational Materials Science}\ }\textbf {\bibinfo
  {volume} {50}},\ \bibinfo {pages} {2075} (\bibinfo {year}
  {2011})}\BibitemShut {NoStop}%
\bibitem [{\citenamefont {Eimre}\ \emph {et~al.}(2015)\citenamefont {Eimre},
  \citenamefont {Parviainen}, \citenamefont {Aabloo}, \citenamefont
  {Djurabekova},\ and\ \citenamefont {Zadin}}]{eimre_application_2015}%
  \BibitemOpen
  \bibfield  {author} {\bibinfo {author} {\bibfnamefont {K.}~\bibnamefont
  {Eimre}}, \bibinfo {author} {\bibfnamefont {S.}~\bibnamefont {Parviainen}},
  \bibinfo {author} {\bibfnamefont {A.}~\bibnamefont {Aabloo}}, \bibinfo
  {author} {\bibfnamefont {F.}~\bibnamefont {Djurabekova}}, \ and\ \bibinfo
  {author} {\bibfnamefont {V.}~\bibnamefont {Zadin}},\ }\href {\doibase
  10.1063/1.4926490} {\bibfield  {journal} {\bibinfo  {journal} {Journal of
  Applied Physics}\ }\textbf {\bibinfo {volume} {118}},\ \bibinfo {pages}
  {033303} (\bibinfo {year} {2015})}\BibitemShut {NoStop}%
\bibitem [{\citenamefont {Charbonnier}\ \emph {et~al.}(1964)\citenamefont
  {Charbonnier}, \citenamefont {Strayer}, \citenamefont {Swanson},\ and\
  \citenamefont {Martin}}]{charbonnier_nottingham_1964}%
  \BibitemOpen
  \bibfield  {author} {\bibinfo {author} {\bibfnamefont {F.~M.}\ \bibnamefont
  {Charbonnier}}, \bibinfo {author} {\bibfnamefont {R.~W.}\ \bibnamefont
  {Strayer}}, \bibinfo {author} {\bibfnamefont {L.~W.}\ \bibnamefont
  {Swanson}}, \ and\ \bibinfo {author} {\bibfnamefont {E.~E.}\ \bibnamefont
  {Martin}},\ }\href {\doibase 10.1103/PhysRevLett.13.397} {\bibfield
  {journal} {\bibinfo  {journal} {Physical Review Letters}\ }\textbf {\bibinfo
  {volume} {13}},\ \bibinfo {pages} {397} (\bibinfo {year} {1964})}\BibitemShut
  {NoStop}%
\bibitem [{\citenamefont {Paulini}\ \emph {et~al.}(1993)\citenamefont
  {Paulini}, \citenamefont {Klein},\ and\ \citenamefont
  {Simon}}]{paulini_thermo-field_1993}%
  \BibitemOpen
  \bibfield  {author} {\bibinfo {author} {\bibfnamefont {J.}~\bibnamefont
  {Paulini}}, \bibinfo {author} {\bibfnamefont {T.}~\bibnamefont {Klein}}, \
  and\ \bibinfo {author} {\bibfnamefont {G.}~\bibnamefont {Simon}},\ }\href
  {http://stacks.iop.org/0022-3727/26/i=8/a=024} {\bibfield  {journal}
  {\bibinfo  {journal} {Journal of Physics D: Applied Physics}\ }\textbf
  {\bibinfo {volume} {26}},\ \bibinfo {pages} {1310} (\bibinfo {year}
  {1993})}\BibitemShut {NoStop}%
\bibitem [{\citenamefont {Franz}\ and\ \citenamefont
  {Wiedemann}(1853)}]{franz_ueber_1853}%
  \BibitemOpen
  \bibfield  {author} {\bibinfo {author} {\bibfnamefont {R.}~\bibnamefont
  {Franz}}\ and\ \bibinfo {author} {\bibfnamefont {G.}~\bibnamefont
  {Wiedemann}},\ }\href {\doibase 10.1002/andp.18531650802} {\bibfield
  {journal} {\bibinfo  {journal} {Annalen der Physik und Chemie}\ }\textbf
  {\bibinfo {volume} {165}},\ \bibinfo {pages} {497} (\bibinfo {year}
  {1853})}\BibitemShut {NoStop}%
\bibitem [{\citenamefont {Schuster}\ \emph {et~al.}(2001)\citenamefont
  {Schuster}, \citenamefont {Vangel},\ and\ \citenamefont
  {Scha}}]{schuster_improved_2001}%
  \BibitemOpen
  \bibfield  {author} {\bibinfo {author} {\bibfnamefont {C.~E.}\ \bibnamefont
  {Schuster}}, \bibinfo {author} {\bibfnamefont {M.~G.}\ \bibnamefont
  {Vangel}}, \ and\ \bibinfo {author} {\bibfnamefont {H.~A.}\ \bibnamefont
  {Scha}},\ }\href {\doibase 10.1016/S0026-2714(00)00227-4} {\bibfield
  {journal} {\bibinfo  {journal} {Microelectronics Reliability}\ ,\ \bibinfo
  {pages} {14}} (\bibinfo {year} {2001})}\BibitemShut {NoStop}%
\bibitem [{\citenamefont {Yarimbiyik}\ \emph {et~al.}(2006)\citenamefont
  {Yarimbiyik}, \citenamefont {Schafft}, \citenamefont {Allen}, \citenamefont
  {Zaghloul},\ and\ \citenamefont {Blackburn}}]{yarimbiyik_modeling_2006}%
  \BibitemOpen
  \bibfield  {author} {\bibinfo {author} {\bibfnamefont {A.~E.}\ \bibnamefont
  {Yarimbiyik}}, \bibinfo {author} {\bibfnamefont {H.~A.}\ \bibnamefont
  {Schafft}}, \bibinfo {author} {\bibfnamefont {R.~A.}\ \bibnamefont {Allen}},
  \bibinfo {author} {\bibfnamefont {M.~E.}\ \bibnamefont {Zaghloul}}, \ and\
  \bibinfo {author} {\bibfnamefont {D.~L.}\ \bibnamefont {Blackburn}},\ }\href
  {\doibase 10.1016/j.microrel.2005.09.004} {\bibfield  {journal} {\bibinfo
  {journal} {Microelectronics Reliability}\ }\textbf {\bibinfo {volume} {46}},\
  \bibinfo {pages} {1050} (\bibinfo {year} {2006})}\BibitemShut {NoStop}%
\bibitem [{\citenamefont {Nath}\ and\ \citenamefont
  {Chopra}(1974)}]{nath_thermal_1974}%
  \BibitemOpen
  \bibfield  {author} {\bibinfo {author} {\bibfnamefont {P.}~\bibnamefont
  {Nath}}\ and\ \bibinfo {author} {\bibfnamefont {K.}~\bibnamefont {Chopra}},\
  }\href {\doibase 10.1016/0040-6090(74)90033-9} {\bibfield  {journal}
  {\bibinfo  {journal} {Thin Solid Films}\ }\textbf {\bibinfo {volume} {20}},\
  \bibinfo {pages} {53} (\bibinfo {year} {1974})}\BibitemShut {NoStop}%
\bibitem [{\citenamefont {Berendsen}\ \emph {et~al.}(1984)\citenamefont
  {Berendsen}, \citenamefont {Postma}, \citenamefont {van Gunsteren},
  \citenamefont {DiNola},\ and\ \citenamefont
  {Haak}}]{berendsen_molecular_1984}%
  \BibitemOpen
  \bibfield  {author} {\bibinfo {author} {\bibfnamefont {H.~J.~C.}\
  \bibnamefont {Berendsen}}, \bibinfo {author} {\bibfnamefont {J.~P.~M.}\
  \bibnamefont {Postma}}, \bibinfo {author} {\bibfnamefont {W.~F.}\
  \bibnamefont {van Gunsteren}}, \bibinfo {author} {\bibfnamefont
  {A.}~\bibnamefont {DiNola}}, \ and\ \bibinfo {author} {\bibfnamefont {J.~R.}\
  \bibnamefont {Haak}},\ }\href {\doibase 10.1063/1.448118} {\bibfield
  {journal} {\bibinfo  {journal} {The Journal of Chemical Physics}\ }\textbf
  {\bibinfo {volume} {81}},\ \bibinfo {pages} {3684} (\bibinfo {year}
  {1984})}\BibitemShut {NoStop}%
\bibitem [{\citenamefont {Child}(1911)}]{child_motion_1911}%
  \BibitemOpen
  \bibfield  {author} {\bibinfo {author} {\bibfnamefont {C.~D.}\ \bibnamefont
  {Child}},\ }\href@noop {} {\bibfield  {journal} {\bibinfo  {journal} {Phys.
  Rev.}\ }\textbf {\bibinfo {volume} {32}},\ \bibinfo {pages} {498} (\bibinfo
  {year} {1911})}\BibitemShut {NoStop}%
\bibitem [{\citenamefont {Barbour}\ \emph {et~al.}(1953)\citenamefont
  {Barbour}, \citenamefont {Dolan}, \citenamefont {Trolan}, \citenamefont
  {Martin},\ and\ \citenamefont {Dyke}}]{barbour_space-charge_1953}%
  \BibitemOpen
  \bibfield  {author} {\bibinfo {author} {\bibfnamefont {J.~P.}\ \bibnamefont
  {Barbour}}, \bibinfo {author} {\bibfnamefont {W.~W.}\ \bibnamefont {Dolan}},
  \bibinfo {author} {\bibfnamefont {J.~K.}\ \bibnamefont {Trolan}}, \bibinfo
  {author} {\bibfnamefont {E.~E.}\ \bibnamefont {Martin}}, \ and\ \bibinfo
  {author} {\bibfnamefont {W.~P.}\ \bibnamefont {Dyke}},\ }\href {\doibase
  10.1103/PhysRev.92.45} {\bibfield  {journal} {\bibinfo  {journal} {Physical
  Review}\ }\textbf {\bibinfo {volume} {92}},\ \bibinfo {pages} {45} (\bibinfo
  {year} {1953})}\BibitemShut {NoStop}%
\bibitem [{\citenamefont {Langmuir}(1913)}]{langmuir_effect_1913}%
  \BibitemOpen
  \bibfield  {author} {\bibinfo {author} {\bibfnamefont {I.}~\bibnamefont
  {Langmuir}},\ }\href {\doibase 10.1103/PhysRev.2.450} {\bibfield  {journal}
  {\bibinfo  {journal} {Physical Review}\ }\textbf {\bibinfo {volume} {2}},\
  \bibinfo {pages} {450} (\bibinfo {year} {1913})}\BibitemShut {NoStop}%
\bibitem [{\citenamefont {Langmuir}(1923)}]{langmuir_effect_1923}%
  \BibitemOpen
  \bibfield  {author} {\bibinfo {author} {\bibfnamefont {I.}~\bibnamefont
  {Langmuir}},\ }\href {\doibase 10.1103/PhysRev.21.419} {\bibfield  {journal}
  {\bibinfo  {journal} {Physical Review}\ }\textbf {\bibinfo {volume} {21}},\
  \bibinfo {pages} {419} (\bibinfo {year} {1923})}\BibitemShut {NoStop}%
\bibitem [{\citenamefont {Kildemo}(2004)}]{kildemo_new_2004}%
  \BibitemOpen
  \bibfield  {author} {\bibinfo {author} {\bibfnamefont {M.}~\bibnamefont
  {Kildemo}},\ }\href {\doibase 10.1016/j.nima.2004.04.230} {\bibfield
  {journal} {\bibinfo  {journal} {Nuclear Instruments and Methods in Physics
  Research Section A: Accelerators, Spectrometers, Detectors and Associated
  Equipment}\ }\textbf {\bibinfo {volume} {530}},\ \bibinfo {pages} {596}
  (\bibinfo {year} {2004})}\BibitemShut {NoStop}%
\bibitem [{\citenamefont {Veske}\ \emph {et~al.}(2016)\citenamefont {Veske},
  \citenamefont {Parviainen}, \citenamefont {Zadin}, \citenamefont {Aabloo},\
  and\ \citenamefont {Djurabekova}}]{veske_electrodynamics-molecular_2016}%
  \BibitemOpen
  \bibfield  {author} {\bibinfo {author} {\bibfnamefont {M.}~\bibnamefont
  {Veske}}, \bibinfo {author} {\bibfnamefont {S.}~\bibnamefont {Parviainen}},
  \bibinfo {author} {\bibfnamefont {V.}~\bibnamefont {Zadin}}, \bibinfo
  {author} {\bibfnamefont {A.}~\bibnamefont {Aabloo}}, \ and\ \bibinfo {author}
  {\bibfnamefont {F.}~\bibnamefont {Djurabekova}},\ }\href {\doibase
  10.1088/0022-3727/49/21/215301} {\bibfield  {journal} {\bibinfo  {journal}
  {Journal of Physics D: Applied Physics}\ }\textbf {\bibinfo {volume} {49}},\
  \bibinfo {pages} {215301} (\bibinfo {year} {2016})}\BibitemShut {NoStop}%
\bibitem [{\citenamefont {Jansson}\ \emph {et~al.}(2016)\citenamefont
  {Jansson}, \citenamefont {Baibuz},\ and\ \citenamefont
  {Djurabekova}}]{jansson_long-term_2016}%
  \BibitemOpen
  \bibfield  {author} {\bibinfo {author} {\bibfnamefont {V.}~\bibnamefont
  {Jansson}}, \bibinfo {author} {\bibfnamefont {E.}~\bibnamefont {Baibuz}}, \
  and\ \bibinfo {author} {\bibfnamefont {F.}~\bibnamefont {Djurabekova}},\
  }\href {\doibase 10.1088/0957-4484/27/26/265708} {\bibfield  {journal}
  {\bibinfo  {journal} {Nanotechnology}\ }\textbf {\bibinfo {volume} {27}},\
  \bibinfo {pages} {265708} (\bibinfo {year} {2016})}\BibitemShut {NoStop}%
\bibitem [{\citenamefont {Nordlund}(1994)}]{nordlund_molecular_1994}%
  \BibitemOpen
  \bibfield  {author} {\bibinfo {author} {\bibfnamefont {K.}~\bibnamefont
  {Nordlund}},\ }\href {\doibase 10.1016/0927-0256(94)00085-Q} {\bibfield
  {journal} {\bibinfo  {journal} {Computational Materials Science}\ }\textbf
  {\bibinfo {volume} {3}},\ \bibinfo {pages} {448} (\bibinfo {year}
  {1994})}\BibitemShut {NoStop}%
\bibitem [{\citenamefont {Nordlund}\ \emph {et~al.}(1998)\citenamefont
  {Nordlund}, \citenamefont {Ghaly}, \citenamefont {Averback}, \citenamefont
  {Caturla}, \citenamefont {de~La~Rubia},\ and\ \citenamefont
  {Tarus}}]{nordlund_defect_1998}%
  \BibitemOpen
  \bibfield  {author} {\bibinfo {author} {\bibfnamefont {K.}~\bibnamefont
  {Nordlund}}, \bibinfo {author} {\bibfnamefont {M.}~\bibnamefont {Ghaly}},
  \bibinfo {author} {\bibfnamefont {R.~S.}\ \bibnamefont {Averback}}, \bibinfo
  {author} {\bibfnamefont {M.}~\bibnamefont {Caturla}}, \bibinfo {author}
  {\bibfnamefont {T.~D.}\ \bibnamefont {de~La~Rubia}}, \ and\ \bibinfo {author}
  {\bibfnamefont {J.}~\bibnamefont {Tarus}},\ }\href
  {http://journals.aps.org/prb/abstract/10.1103/PhysRevB.57.7556} {\bibfield
  {journal} {\bibinfo  {journal} {Physical Review B}\ }\textbf {\bibinfo
  {volume} {57}},\ \bibinfo {pages} {7556} (\bibinfo {year}
  {1998})}\BibitemShut {NoStop}%
\bibitem [{\citenamefont {Ghaly}\ \emph {et~al.}(1999)\citenamefont {Ghaly},
  \citenamefont {Nordlund},\ and\ \citenamefont
  {Averback}}]{ghaly_molecular_1999}%
  \BibitemOpen
  \bibfield  {author} {\bibinfo {author} {\bibfnamefont {M.}~\bibnamefont
  {Ghaly}}, \bibinfo {author} {\bibfnamefont {K.}~\bibnamefont {Nordlund}}, \
  and\ \bibinfo {author} {\bibfnamefont {R.~S.}\ \bibnamefont {Averback}},\
  }\href {\doibase 10.1080/01418619908210332} {\bibfield  {journal} {\bibinfo
  {journal} {Philosophical Magazine A}\ }\textbf {\bibinfo {volume} {79}},\
  \bibinfo {pages} {795} (\bibinfo {year} {1999})}\BibitemShut {NoStop}%
\bibitem [{\citenamefont {Bangerth}\ \emph {et~al.}(2007)\citenamefont
  {Bangerth}, \citenamefont {Hartmann},\ and\ \citenamefont
  {Kanschat}}]{bangerth_deal.ii_2007}%
  \BibitemOpen
  \bibfield  {author} {\bibinfo {author} {\bibfnamefont {W.}~\bibnamefont
  {Bangerth}}, \bibinfo {author} {\bibfnamefont {R.}~\bibnamefont {Hartmann}},
  \ and\ \bibinfo {author} {\bibfnamefont {G.}~\bibnamefont {Kanschat}},\
  }\href {\doibase 10.1145/1268776.1268779} {\bibfield  {journal} {\bibinfo
  {journal} {ACM Transactions on Mathematical Software}\ }\textbf {\bibinfo
  {volume} {33}},\ \bibinfo {pages} {24/1} (\bibinfo {year}
  {2007})}\BibitemShut {NoStop}%
\bibitem [{\citenamefont {Mishin}\ \emph {et~al.}(2001)\citenamefont {Mishin},
  \citenamefont {Mehl}, \citenamefont {Papaconstantopoulos}, \citenamefont
  {Voter},\ and\ \citenamefont {Kress}}]{mishin_structural_2001}%
  \BibitemOpen
  \bibfield  {author} {\bibinfo {author} {\bibfnamefont {Y.}~\bibnamefont
  {Mishin}}, \bibinfo {author} {\bibfnamefont {M.}~\bibnamefont {Mehl}},
  \bibinfo {author} {\bibfnamefont {D.}~\bibnamefont {Papaconstantopoulos}},
  \bibinfo {author} {\bibfnamefont {A.}~\bibnamefont {Voter}}, \ and\ \bibinfo
  {author} {\bibfnamefont {J.}~\bibnamefont {Kress}},\ }\href {\doibase
  10.1103/PhysRevB.63.224106} {\bibfield  {journal} {\bibinfo  {journal}
  {Physical Review B}\ }\textbf {\bibinfo {volume} {63}},\ \bibinfo {pages}
  {224106} (\bibinfo {year} {2001})}\BibitemShut {NoStop}%
\bibitem [{\citenamefont {Descoeudres}\ \emph {et~al.}(2009)\citenamefont
  {Descoeudres}, \citenamefont {Levinsen}, \citenamefont {Calatroni},
  \citenamefont {Taborelli},\ and\ \citenamefont
  {Wuensch}}]{descoeudres_investigation_2009}%
  \BibitemOpen
  \bibfield  {author} {\bibinfo {author} {\bibfnamefont {A.}~\bibnamefont
  {Descoeudres}}, \bibinfo {author} {\bibfnamefont {Y.}~\bibnamefont
  {Levinsen}}, \bibinfo {author} {\bibfnamefont {S.}~\bibnamefont {Calatroni}},
  \bibinfo {author} {\bibfnamefont {M.}~\bibnamefont {Taborelli}}, \ and\
  \bibinfo {author} {\bibfnamefont {W.}~\bibnamefont {Wuensch}},\ }\href
  {\doibase 10.1103/PhysRevSTAB.12.092001} {\bibfield  {journal} {\bibinfo
  {journal} {Physical Review Special Topics - Accelerators and Beams}\ }\textbf
  {\bibinfo {volume} {12}},\ \bibinfo {pages} {092001} (\bibinfo {year}
  {2009})}\BibitemShut {NoStop}%
\bibitem [{\citenamefont {Lewis}\ \emph {et~al.}(2004)\citenamefont {Lewis},
  \citenamefont {Nithiarasu},\ and\ \citenamefont
  {Seetharamu}}]{lewis_fundamentals_2004}%
  \BibitemOpen
  \bibfield  {author} {\bibinfo {author} {\bibfnamefont {R.~W.}\ \bibnamefont
  {Lewis}}, \bibinfo {author} {\bibfnamefont {P.}~\bibnamefont {Nithiarasu}}, \
  and\ \bibinfo {author} {\bibfnamefont {K.~N.}\ \bibnamefont {Seetharamu}},\
  }\href@noop {} {\emph {\bibinfo {title} {Fundamentals of the finite element
  method for heat and fluid flow}}}\ (\bibinfo  {publisher} {Wiley},\ \bibinfo
  {year} {2004})\BibitemShut {NoStop}%
\bibitem [{\citenamefont {Vince}(2014)}]{vince_barycentric_2014}%
  \BibitemOpen
  \bibfield  {author} {\bibinfo {author} {\bibfnamefont {J.}~\bibnamefont
  {Vince}},\ }in\ \href@noop {} {\emph {\bibinfo {booktitle} {Mathematics for
  {Computer} {Graphics}}}}\ (\bibinfo  {publisher} {Springer},\ \bibinfo {year}
  {2014})\ \bibinfo {edition} {4th}\ ed.,\ pp.\ \bibinfo {pages}
  {203--229}\BibitemShut {NoStop}%
\bibitem [{\citenamefont {Krall}\ and\ \citenamefont
  {Trivelpiece}(1973)}]{krall_principles_1973}%
  \BibitemOpen
  \bibfield  {author} {\bibinfo {author} {\bibfnamefont {N.~A.}\ \bibnamefont
  {Krall}}\ and\ \bibinfo {author} {\bibfnamefont {A.~W.}\ \bibnamefont
  {Trivelpiece}},\ }\href@noop {} {\emph {\bibinfo {title} {Principles of
  {Plasma} {Physics}}}},\ \bibinfo {edition} {1st}\ ed.\ (\bibinfo  {publisher}
  {McGraw-Hill},\ \bibinfo {year} {1973})\BibitemShut {NoStop}%
\end{thebibliography}%

\end{document}